\documentclass{article}
\usepackage{float}
\usepackage{subfigure}
\usepackage{float}
\usepackage{amsmath}
\usepackage{color}
\usepackage{graphicx}
\usepackage{epsfig}
\usepackage{algorithm}
\usepackage[noend]{algorithmic}
\usepackage[left=2.5cm,top=2.5cm,right=2.5cm,bottom=2cm,nohead,footskip=0.5cm]{geometry}
\usepackage{appendix}

\begin{document}
\title{Inferring Neuronal Network Connectivity from Spike Data: A Temporal 
Data Mining Approach}
\author{
\begin{tabular}[t]{ccc} 
Debprakash Patnaik\footnote{Current address: Computer Science Department, Virginia Tech,
Blacksburg, VA 24060, USA}  & P. S. Sastry & K. P. Unnikrishnan\\ 
debprakash@gmail.com & sastry@ee.iisc.ernet.in & k.unnikrishnan@gm.com\\
Dept. of Electrical Engg. & Dept. of Electrical Engg. & General Motors R\&D Center\\
Indian Institute of Science & Indian Institute of Science & Waren, MI 48090-9055, USA\\
Bangalore, 560012 India & Bangalore, 560012 India & \\
\end{tabular}}
\date{}
\maketitle

\begin{abstract}
Understanding the functioning of a neural system in terms of its underlying circuitry is an 
important problem in neuroscience.  Recent developments in electrophysiology and imaging  
allow one to simultaneously record activities of hundreds of neurons. Inferring the underlying 
neuronal connectivity patterns from such multi-neuronal spike train data streams is a 
challenging statistical and computational problem. 
This task involves finding significant temporal patterns from vast 
amounts of symbolic time series data. In this paper we show that 
the frequent episode mining methods from the field of 
temporal data mining can be very useful in this context. In the frequent episode discovery 
framework, the data is viewed as 
 a sequence of events, each of which is characterized by an event type and its time of 
occurrence and episodes are certain types of temporal patterns in such data. 
Here we show that, using the set of discovered frequent episodes from multi-neuronal data, 
one can infer different 
types of connectivity patterns in the neural system that generated it. For this purpose, we 
introduce the notion of mining for frequent episodes under certain temporal constraints; the 
structure of these temporal constraints is motivated by the application. 
We present algorithms 
for discovering serial and parallel episodes under these temporal constraints. 
Through extensive simulation studies we 
 demonstrate that these methods are useful for unearthing 
patterns of neuronal network connectivity.  
\end{abstract}

\small{\quotation{
}}

\section{Introduction}
Over the last couple of decades, biology has thrown up many interesting and challenging 
computational problems. For example, the problem of understanding genome data and protein 
function has motivated development of many computational and statistical 
techniques leading to the creation of the interdisciplinary area of Bioinformatics. One 
of the main driving forces in this case is the availability of large amounts of data, 
from gene or protein sequencing experiments, and the consequent need for 
efficient techniques to analyze the data to arrive at reasonable and useful inferences. 
To solve these computational problems, some techniques developed in other contexts 
(e.g., Hidden Markov Models, Dynamic Programming) 
have proved to be quite suitable, after some modifications. 

In this paper, we focus on an equally challenging computational problem in another sub area 
of biology, namely neuroscience. We look at the problem 
of analyzing multi-neuronal spike train data and suggest that certain techniques from 
the field of Temporal Data mining are attractive here. 

Neurons form the basic computing elements of brain and hence, gaining an understanding 
of the coordinated behavior of groups of neurons (at different levels of organization) is 
essential for gaining a principled understanding of brain function. Thus, one of the 
important problems in neuroscience is that of understanding the functioning of a   
neural tissue in terms of interactions among its neurons. Many neurons communicate 
with each other through characteristic electric pulses called action potentials or spikes. 
Hence one can study the activity 
of a specific neural tissue by gathering data in the form of sequences of 
action potentials or spikes generated 
by each of a group of potentially interconnected neurons. Such data is known as 
multi-neuronal spike train data. (See section~\ref{sec:mea} for more details). 
 
Over the past twenty years or so, increasingly better methods are becoming available 
for simultaneously recording the activities of many neurons. 
By using techniques such as micro electrode arrays, imaging of
currents, voltages, and ionic concentrations etc., spike data can be recorded
simultaneously from hundreds of neurons 
\cite{Ikegaya2004,ludwig-et-al-2005,vetter-et-al-2004,Potter2006}.  Vast amounts of 
such data is now routinely gathered from different neuronal systems. 
For example, in \cite{Potter2006} the authors describe experiments where tens of 
cortical cultures are maintained for over five weeks and on each day the spiking 
activities of neurons (both with and without external stimulation) in each culture 
are recorded for tens of minutes. Each recording session contains data with tens of 
thousands of spikes. 
Such multi-neuronal spike train data can now be obtained 
{\em in vitro} from neuronal cultures or {\em in vivo} from 
brain slices, awake behaving animals, and even humans. 
Such  spike train data is a mixture of the stochastic spiking 
activities of individual neurons as well as correlated 
spiking activity due to interactions or 
connections among neurons.

Availability of such data motivates development of efficient techniques for 
analyzing the spike train data \cite{Brown2004}. 
The computational challenge is to make reasonable inferences regarding  the 
connectivity information or the microcircuits present in the neuronal tissue.  
The grand objective is to come out with a host of data 
processing and analysis techniques that would enable reliable inference of the 
underlying functional connectivity patterns which characterize the 
microcircuits in the neuronal systems \cite{Brown2004}. 
Like in the field of Bioinformatics, this endeavor also entails an 
interdisciplinary approach and we may call such an approach as Neuroinformatics. 
The main objective of this paper is to show that techniques from Temporal 
Data mining could offer novel and useful points of view for tackling some of the 
issues involved in analyzing spike train data.  

Temporal data mining is concerned with analyzing symbolic data streams with temporal 
dependencies to discover `interesting' temporal 
patterns \cite{Srivats-survey2005,Morchen2007,UU2001,UUH-2004,URSU-2006}. 
Temporal data mining
differs from classical time series analysis in the kind of information
that one seeks to discover. The exact model parameters (e.g. coefficients
of an ARMA model or the weights of a recurrent neural network) are
of little interest. Unearthing interesting trends or patterns
in the data (which are much more readily interpretable by the data 
owner), is of primary interest. 
Here we focus on  the frequent 
episodes framework of temporal data mining \cite{Mannila1997,Srivats2005,Srivats2007}. 
In this framework, we view the input data as a sequence of {\em events} with each 
event characterized by an {\em event type} 
 and a {\em time of occurrence}. The patterns to be discovered 
are called {\em episodes}. (See Sec.~\ref{sec:epi} for more details).   This framework 
is found useful in many engineering applications. For example,  
the sequence of fault records logged in a manufacturing process can be viewed as a 
data stream of events.  Then the episodes capture faults with temporal correlations and 
hence frequent episode discovery can be useful in root cause diagnosis 
\cite{Srivats2007}. 

The multi-neuronal spike train data can also be viewed as 
a sequential or time-ordered data stream of events where each event is a spike at
a particular time and
 the event type is the neuron that generated the spike.
 Since functionally
interconnected neurons tend to fire in certain precise patterns, discovering
frequent episodes in such temporal data can help understand the underlying
neural circuitry.
In this paper we present some novel techniques for frequent episode discovery 
and show their utility for the analysis of multi-neuronal spike train data.

Most of the currently proposed methods for analyzing spike train data
rely on quantities that can be
computed through cross correlations among spike trains (time shifted with respect to
one another) \cite{Brown2004}.
Most of these methods are not computationally 
efficient for {\em discovering}
temporal patterns that involve more than a few neurons.
(See Sec~\ref{sec:mea} for a review of spike data analysis). 
Here we show that the  temporal data mining approach of frequent episode 
discovery is very effective for coming up with reasonable hypotheses regarding the
connectivity pattern. Though data mining has been successful in unearthing interesting
patterns in large databases in many engineering applications, to the best of our
knowledge, this is the first time a data mining technique is explored for spike train
data analysis. In our opinion, there are mainly two reasons why a data mining approach
is attractive for this problem. Firstly, these techniques are known to be efficient
in tackling the combinatorial explosion while looking for patterns in large data sets.
Thus, we would be able to efficiently discover patterns involving many neurons also.
Secondly, these techniques are essentially model independent in the sense that the
discovery algorithms do not need to make many assumptions about the kind of interactions
among the neurons in terms of, e.g., parameterized classes of models. 
The empirical results we present here illustrate both these aspects.

The rest of the paper is organized as follows. 
We explain the problem of analyzing multi-neuronal spike train data in 
Section~\ref{sec:mea}. We then present a 
brief overview of the frequent episodes framework in Section~\ref{sec:epi}. 
We also introduce a framework of imposing some temporal constraints on the episode 
occurrences which will be needed in the multi-neuronal data analysis. 
We end this section with a discussion that  explains  
how one can use methods of serial and parallel episode discovery, under temporal 
constraints, to discover many patterns of interest in the spike train data. 
While there are many efficient algorithms for discovering frequent episodes, 
none of these can directly handle the kind of temporal constraints that are 
needed in this application. We present some novel 
 algorithms for discovering frequent episodes under temporal constraints 
 in Section~\ref{sec:algos}. We present some simulation results to 
illustrate our method of discovering connection patterns in neuronal networks 
in Section~\ref{sec:res}. For this we have built a simulator for generating spike train 
data by modeling each neuron as an inhomogeneous Poisson process whose firing rate 
changes as a function of input spikes it receives from other neurons. We choose 
two different functional forms for changing the firing rate as a function of input 
received by a neuron so as to illustrate that our discovery algorithms and their 
effectiveness in inferring connectivity patterns, are not dependent on any specific 
mechanism of interactions among neurons. The simulator 
generates fairly realistic spike data and it allows us to validate the algorithms 
presented here. We also discuss some results on data from cortical cultures. 
Finally, we conclude the paper in Section~\ref{sec:conc}.

\section{Multi-neuronal Spike Train Data and Its Analysis}
\label{sec:mea}

The brain or the nervous system consists essentially of a vast network of neurons. The 
neuron may be regarded as the basic computing element in the nervous system. Each 
neuron is connected to many others through what are known as synapses. Synapses, 
through electric and chemical means, allow neurons to signal to each other in the sense 
that the output of one neuron can become input to another through the synapse that 
connects them. In a good majority of all neurons, the output of a neuron is in the 
form of what is called an action potential. An action potential is an electrical signal 
of a short duration (typically less than 1 ms) with a characteristic shape. For 
most purposes of analysis, this can be regarded as a short pulse and hence is also 
referred to as a spike. After generating an action potential, a neuron can not 
immediately generate another spike because it needs some regenerative time. This 
time period is called refractory period and in many cases it is in the range 
of 1 milli second (ms). 
Over short durations of time the spiking activity of a neuron can be well modeled 
by a Poisson process whose rate depends on the current state of the neuron. 
The spikes output by one neuron reach the input terminals 
of other neurons through the synapses that interconnect them. Each neuron has many 
such synapses and based on the amount of input it receives like this, it may then 
fire an action potential or a spike. The functioning of the nervous system is 
essentially due to this coordinated activity of many neurons. (See, e.g., 
\cite{dayan-book} for a good exposition). The system is stochastic and 
neurons would also be spiking randomly. The signal transmission through synapses 
takes some time and thus there are characteristic delays associated with each 
synapse. Also different synapses may have different efficacies in effecting 
spikes from the receiving neurons. 

Experimental studies for understanding brain function span a wide range of organizational 
levels. At one end are studies aimed at understanding the functioning of single neurons 
through electro physiological recordings while at the other end, using techniques such as 
fMRI, one studies interactions among large brain regions. Our interest in this paper is in 
experimental techniques at an intermediate level of organization where one is interested 
in understanding how groups of a few hundred neurons 
act in a coordinated manner to generate specific functions. For this, as mentioned earlier, 
one obtains simultaneous recordings of  the spikes generated by a group of interacting 
neurons. By simultaneous recording we mean that the times of spikes of all neurons are 
referenced with respect to a common time origin and hence the data is suitable for studying 
temporal interactions among the neurons. 

There has been a lot of work on experiments for 
 simultaneously recording the activities of hundreds of neurons for gaining 
a better understanding of the functional interactions among neurons in a neural tissue 
\cite{Abeles1988,Meister1994,Meister1996,Nicolelis2003,Gerstein2004,WiseAnderson2004,
Ikegaya2004,Hosoya2005,Potter2006}.  
The recording techniques fall into three broad categories. 
In the first category are recordings 
from cultured neurons or brain slices using Micro electrode arrays (MEAs). 
A typical MEA setup for this consists of $8\times 8$ grid of 64 electrodes with inter-electrode
spacing of about 100 microns. This allows stimulation of the neural tissue and recording 
of the resulting spikes using the same set of electrodes \cite{Potter2006,mea-slice-2002}.  
In the second category are recordings from intact animals using MEAs and other probes  
\cite{vetter-et-al-2004,ludwig-et-al-2005,Nicolelis2003,WiseAnderson2004}. 
In the third category are imaging techniques
using voltage sensitive dyes and indicators for ions such as Ca++ 
\cite{Ikegaya2004,sasaki2007}. 
 One of the most exciting recent developments is the incorporation of ion-selective 
pores into neurons of behaving animals \cite{Zhang-et-al-2007,HB2007}. 
This allows simultaneous stimulation and
recording with milli-second precision using light at various wavelengths.
All these technologies now allow for gathering of vast amounts of data, using which one wishes
to study connectivity patterns and microcircuits in neural systems.

In this paper our interest is in techniques for analyzing the data that is in the 
form of spike trains. To obtain such spike trains from the recorded data, one needs 
a lot of signal processing and data preprocessing techniques. For example, in MEA 
experiments, the raw data is in the form of voltage or current signals from each 
of the electrodes, recorded at a suitably high sampling rate. By employing appropriate 
signal processing techniques one has to first reliably locate all spiking events. Even 
after this, what we have are spike events in each channel or each electrode. Since 
the micro electrode array is regular while the neuronal tissue is not, each electrode 
may be picking up signals from many neurons with different efficacies. If we want 
the final data as spike events generated by individual neurons then we have to 
do what is called {\em spike sorting}. Many techniques have been suggested for 
processing the raw signals to obtain spikes and to do spike sorting and there is 
need for better algorithms for these problems. (See, e.g., 
\cite{Brown2004} and references therein). Here we will not review any of these 
techniques because our interest is in analysis methods that look at spike trains to 
infer connectivity patterns.

The field of multi-neuronal data analysis has a
long history, beginning with the work of Gerstien and his
colleagues \cite{GersteinPerkel1969,GersteinPerkel1972,Abeles1988,gersteinGravity1985}.
A major goal of such neural data analysis is to characterize how neurons
that are part of an ensemble interact with each other. 
Statistical analysis of spike train data was pioneered by 
Brillinger \cite{Brillinger1992} and the
recent review by Kass, Ventura, and Brown \cite{Kass2005} addresses 
all the statistical issues in this area.
The review by Brown, Kass, and Mitra \cite{Brown2004} summarizes three 
decades of methodology 
development in this area and eloquently lays out future challenges.
Many of the current methods essentially use information obtained from 
cross correlation among spike trains that are shifted in time with respect 
to one another \cite{Brown2004}. For example, one can compute what is called a 
joint peri-stimulus time histogram (JPSTH) which is a two dimensional histogram 
that displays the joint spike count of two neurons at different time lags (for 
a specific binning on the time axis). There are other methods based on analyzing 
time-shifted spike trains for detecting repeated patterns of firing of a few neurons 
with constant time lags \cite{Abeles1988,Abeles2001,Tetko2001}. Given  some 
specific patterns there are methods to look for matches in the spike train data and 
assess the statistical significance \cite{Lee2004}. For assessing the statistical 
significance of the detected patterns, one generally employs a null hypothesis 
that the spike trains are {\em iid} Bernoulli processes or uses some resampling 
methods of generating surrogate data streams to assess significance 
empirically \cite{Gerstein2004,Abeles1988,Lee2004}. Most of these techniques 
are not efficient for detecting patterns involving more than four or five  
neurons. Another approach is to 
employ dimensionality reduction techniques such as PCA and study the data in 
some appropriate low dimensional feature space \cite{sasaki2007}. Bayesian 
model estimation techniques have also been used to infer parameters of an 
assumed statistical model of interaction among neurons \cite{Rigat-bayes-2006}.

The availability of vast amounts data means that developing efficient methods
to analyze neuronal spike trains is a challenging task of immediate utility
in this area \cite{Brown2004}. To quote Brown, Kass, and Mitra \cite{Brown2004}, 
``{\em Simultaneous recording
of multiple spike trains from several neurons offers a window into how neurons work in
concert to generate specific brain functions. Without
substantial methodology research in the future, our ability to understand this function will be
significantly hampered because current methods fall short of what is ultimately 
required for the analysis of multiple spike train data.}''

The objective of analyzing spike train data is to finally be able to infer 
the microcircuits and relate the functioning of neural systems to specific 
 coordinated activities or characteristic firing 
patterns \cite{Bialek1991,Abeles-book}. In this paper   
we address the simpler problem of inferring some useful temporal patterns in 
the spike trains that help understand the connectivity structure.   

The characteristics of patterns that one is interested in  can be roughly grouped into
Synchrony, Order, Synfire chains, and Synfire braids. Synchronous firing by a group of
neurons is interesting because it can be an efficient way to transmit information
\cite{Hosoya2005,Riehle2000,Schnitzer2003}.
Ordered or sequential firing of neurons where the time-delays between
firing of successive neurons are fairly constant denote a chain of triggering
events and unearthing such relations between neurons can thus reveal 
underlying functional connectivity \cite{Abeles2001}.
 If neuron A is functionally connected to
neuron B, it influences the firing of neuron B. If this is an excitatory connection
(with or without a delay), then, if A fires, B is likely to fire soon after that. Hence,
discovering the order of neuronal firings (along with the delays involved) 
can help decipher the functional connectivity. Memory traces are probably embedded 
in such sequential activations of neurons or neuronal groups.
Signals of this form have recently been found in groups of hippocampal neurons by
Lee and Wilson \cite{Lee2002}. 
An ordered chain of firings of neuronal groups (rather than single neurons) is sometimes
called a Synfire chain and is believed to be an 
important microcircuit \cite{Abeles-book,Ikegaya2004}.
A synfire chain can be thought of as a compound pattern involving both synchrony and
order. Synfire braids, also called Polychronous chains,  are
generalizations of Synfire chains. Here, a group of neurons are
activated in fairly precise temporal relationships and automatic discovery of these
polychronous circuits are considered a very difficult task \cite{Izh2006}.

Discovering such temporal patterns in spike trains
amounts to unearthing groups of neurons
that fire in some kind of coordinated fashion. As already mentioned, in most of the
currently available methods, the curse of dimensionality forces the analysis
to be confined to a few variables at a time.
For the same reason, it is often very difficult
to {\em discover} all patterns of a particular kind. Thus, many of the
available  algorithms are for
counting occurrences of specific list of patterns. In the next section we explain the idea
of frequent episodes and show that this data mining viewpoint
gives us a unified algorithmic
scheme for {\em discovering} many types of interesting patterns in spike train data.


\section{Frequent Episode Discovery}
\label{sec:epi}

Frequent episode discovery framework was proposed by Mannila et.al. 
\cite{Mannila1997} in the context analyzing alarm sequences in a communication 
network. Laxman et.al. \cite{Srivats2005} introduced the notion of 
non-overlapped occurrences as episode frequency and proposed  
efficient counting algorithms. We first give brief overview of this framework.

The data to be analyzed is a sequence of events 
denoted by $\langle(E_{1},t_{1}),(E_{2},t_{2}),\ldots\rangle$ where $E_{i}$ 
represents an \textit{event type} and $t_{i}$ the \textit{time of occurrence} of 
the $i^{th}$ event. $E_i$'s are drawn from a finite set of event types, $\zeta$. 
The sequence is ordered with respect to time of occurrences of the events so 
that, $t_i\le t_{i+1}$, for all $i=1,2,\ldots$. The following is an 
example event sequence containing 9 events with 5 event types.
\begin{equation}
\langle(A,1),(B,3),(D,4),(C,6),(B,8),(A,10),(E,14),(B,15),(C,18)\rangle
\label{eq:data-seq}
\end{equation}

In multi-neuron data, a spike event has the label of the neuron (or the electrode 
number when we consider  multi-electrode array recordings without the  
spike sorting step) which 
generated the spike as its event type  and has the associated time of occurrence. 
The neurons in the ensemble under observation fire action potentials at 
different times, that is, generate spike events. All these spike events are 
strung together, in time order, to give a single long data sequence as needed for 
frequent episode discovery.

The general temporal patterns that we wish to discover in this 
framework are called episodes. 
 In this paper we shall deal with two types of 
episodes: \textit{Serial} and \textit{Parallel}.

Formally, an episode $\alpha$ is a triple 
$(V_{\alpha},\leq_{\alpha},g_{\alpha})$, where $V_{\alpha}$ is a set of nodes, 
$\leq_{\alpha}$ is a partial order on $V_{\alpha}$, and 
$g_{\alpha}:V_{\alpha}\rightarrow \zeta$, is a mapping 
associating each node with an event type. 
The size of $\alpha$, denoted as $|\alpha|$, is 
$|V_{\alpha}|$ (i.e. the number of nodes in $V_{\alpha}$). Episode $\alpha$ is a 
parallel episode if the partial order $\leq_{\alpha}$ is a null set. It is a 
serial episode if the relation $\leq_{\alpha}$ is a total order. 
A non-empty partial order which is neither a total order nor a null set  
corresponds to a general episode. 
An episode is said to occur in an event sequence if there are events
in the data sequence with the same time ordering as specified by the
episode.

A \textit{serial episode} is an ordered tuple of event types. For example, 
$(A\rightarrow B\rightarrow C)$ is a 3-node serial episode. The arrows in this 
notation indicate the order of the events. Such an episode is said to 
\textit{occur} in an event sequence if there are  corresponding events in the 
prescribed order. 
In contrast, a \textit{parallel episode} is similar to an unordered set of items. It 
does not require any specific ordering of the events. We denote a 
3-node parallel episode with event types $A$, $B$ and $C$, as $(ABC)$. An 
occurrence of $(ABC)$ can have the events in any order in the sequence. 
In sequence (\ref{eq:data-seq}), the events 
\{${(A,1),(B,3),(C,6)}$\} constitute an occurrence of the 
serial episode $(A\rightarrow B \rightarrow C)$ while 
the events \{$(B,3), (C,6), (A,10)$\} do not. However, both these 
sets of events constitute occurrences of the parallel episode $(ABC)$. 

We note here that occurrence of an episode (of either type) does not 
require the associated event types to occur consecutively;  
there can be other intervening events between them. 
 In the multi-neuronal data, if neuron $A$ makes 
neuron $B$ to fire, then, we expect to see $B$ following $A$ often. However, in 
different occurrences of such a substring, there may be different number of 
other spikes between $A$ and $B$ because many other neurons may also be spiking 
simultaneously. Thus, the episode structure allows us to unearth patterns  
in the presence of such noise in spike data.

{\em Subepisode}: 
An episode $\beta$ is a sub-episode of episode $\alpha$ if all event types of 
$\beta$ are in $\alpha$ and if partial order among the event types of $\beta$ is 
same as that for the corresponding event types in $\alpha$. For example 
$(A\rightarrow B)$, $(A\rightarrow C)$, and $(B\rightarrow C)$ are 2-node 
sub-episodes of the 3-node episode $(A\rightarrow B\rightarrow C)$, while 
$(B\rightarrow A)$ is not. In case of parallel episodes, there is no ordering 
requirement. Hence every subset of the set of event types of an episode is a 
subepisode. It is to be noted here that occurrence of an episode implies 
occurrence of all its subepisodes.


{\em Frequency of episodes}:
A frequent episode is one whose frequency exceeds a user specified threshold.
The frequency of an episode can be defined in many ways 
\cite{Mannila1997,Srivats2005,Srivats2006}. 
It is intended
to capture some measure of how often an episode occurs in an event
sequence. One chooses a measure of frequency so that frequent episode discovery is
computationally efficient and, at the same time, higher frequency would imply that
an episode is occurring often.
Here, we use the \textit{non-overlapped occurrence} count
 proposed in \cite{Srivats2005}  as the frequency. 

Two occurrences of an episode $\alpha$ are 
said to be \textit{non-overlapped} if no event associated with one occurrence appears in 
between the events associated with the other. 
A collection of 
occurrences of $\alpha$ is said to be non-overlapped if every pair of occurrence 
in it is non-overlapped. The corresponding frequency for episode $\alpha$ is 
defined as the cardinality of the largest set of non-overlapped occurrences of 
$\alpha$ in the given event sequence. (See \cite{Srivats2005} for more discussion). 
This definition of frequency results in very efficient 
counting algorithms with some interesting theoretical 
properties \cite{Srivats2005,Srivats-kdd07}. 
In the context of our application, counting non-overlapped occurrences is 
natural because we would then be looking at causative chains that happen at 
different times again and again.

\subsection{Temporal Constraints}
As stated earlier, while analyzing neuronal spike data, it is useful to consider
methods,
where, while counting the frequency, we include only those occurrences which
satisfy some additional temporal constraints. In a general data mining task, constraints 
provide the user with an ability to focus the pattern discovery task toward 
patterns that are more useful or interesting in the applications. There has been a 
lot of research work aimed at developing efficient algorithms for pattern discovery 
under constraints in the context of frequent itemset mining. (See \cite{Bonchi2007} for 
a very good exposition on the state of art in this area). However, there are no 
general techniques for incorporating temporal constraints in the context of frequent episode 
discovery from event sequences. 
(By temporal constraints we mean constraints that are based on 
the times of occurrences of events).  
In this paper we extend the available frequent 
episode discovery algorithms to take care of some temporal constraints which are useful 
in our application. 
We mainly consider two types of
such constraints: episode expiry time and inter-event time constraints.

Given an episode occurrence (that is, a set of events in the data stream that 
constitute an occurrence of the episode), we call the largest time difference 
between any two events constituting the occurrence as the span of the occurrence. 
For serial episodes, where events constituting an occurrence have to be in a prescribed 
order, the span would be the difference between times of the first 
and last events of the episode in an occurrence. Even for parallel episodes, where 
events can occur in any order, the span as defined above is meaningful. For example, 
consider the occurrences of the parallel episode $(ABC)$ in the example sequence 
(\ref{eq:data-seq}). For the occurrence constituted by the set of events 
$\{(A,1)), \: (C,6), \: (B,8) \}$ the span is 7 units because that is the largest 
time difference between any two events in the occurrence. 
The episode expiry time constraint 
requires that we count only those occurrences whose span is less than a (user-specified) 
time $T_X$. In the algorithm in \cite{Mannila1997}, the window width essentially 
implements an upper bound on the span of occurrences. An efficient algorithm for 
counting non-overlapping occurrences of serial episodes that satisfy an expiry time 
constraint is available in \cite{Srivats2006}. 

The inter-event time constraint, which is meaningful only 
for serial episodes, is specified by giving an interval of  
the form $(T_{low}, T_{high}]$ and requires that the difference between the times of 
 every pair of successive events in any occurrence of a serial episode 
should be in this interval.  In a generalized form of this constraint, we may have 
different time intervals for different pairs of events in each serial episode. 

In the next subsection we explain why these temporal constraints are useful while 
looking for patterns in multi-neuronal spike data. 
While these temporal constraints are motivated by our application, these are 
fairly general and would be useful in many other applications of frequent 
episode discovery. 

In Section~\ref{sec:algos}, we present a new algorithm for 
counting non-overlapped occurrences of serial episodes with such inter-event time 
constraints. We present a general version of the 
 algorithm which can actually choose the most suitable interval constraint 
(by searching over a set of intervals provided by the user) for each consecutive pair of 
events in the discovered frequent episodes. We call this  as discovery of episodes 
under  generalized inter-event time constraints. This algorithm is easily specialized 
for the case where a single interval is given as the inter-event constraint for 
each pair of consecutive nodes.

\subsection{Episodes as patterns in neuronal spike data}

\begin{figure}[!htb]
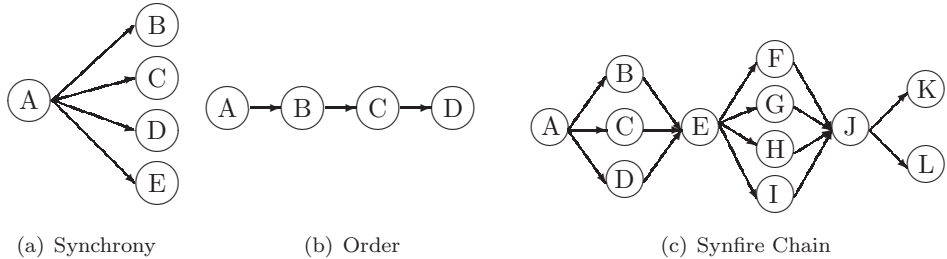

\centering
\subfigure[Synchrony]{\input{parallel-structure.pic}}    
\subfigure[Order]{\input{serial-structure.pic}}
\subfigure[Synfire Chain]{\input{synfire-structure.pic}}
\caption{Examples of neuronal connection structures that can result in different 
patterns in the spike trains: 
 (a). Neurons B,C,D,E may fire synchronously, (b). simple 
circuit that generates firing of A, B, C, D in order, (c). A synfire chain pattern where 
different groups of synchronously firing neurons obey a serial order.}
\label{fig:episode-structures}
\end{figure}

The analysis requirements of spike train data
are met very well by the frequent episodes framework. Serial
and parallel episodes with appropriate temporal
constraints can capture many patterns of interest in 
multi-neuronal data. Fig.~\ref{fig:episode-structures} shows some
possibilities of neuronal interconnections that may give rise to
different patterns in spike data.

As stated earlier, one of the patterns of interest is
Synchrony or co-spiking activity in which groups of neurons
fire synchronously. This kind of synchrony may not be precise. 
Allowing for some amount of variability, co-spiking
activity requires that all neurons must fire within a small interval
of time of each other (in any order) for them to be grouped together.
Such synchronous firing patterns may be generated using the structure
as shown in Fig.~\ref{fig:episode-structures}(a). Here, neurons $B,C,D$ and $E$ fire 
synchronously because all of them are triggered by the same neuron. (Here we 
are assuming that all the synapses are of the same type and hence have the 
same average delay).  There may, however, 
be some variability in the order of their firing and the times at which the spikes 
occur because the delay time through a synapse is often random. 
Such patterns of Synchrony can be discovered 
by looking for frequent parallel episodes which satisfy an expiry time constraint.
For example, we can choose the expiry time to be less than a typical
synaptic delay. This would ensure that spikes from two neurons connected through 
a synapse would not constitute an occurrence of this parallel episode which 
is what is needed when looking for synchrony.  
The expiry time here controls the amount of variability allowed
for declaring a grouped activity as synchronous.

Another pattern in spike data is ordered firings.
A simple mechanism that can generate ordered firing sequences is
shown in Fig.~\ref{fig:episode-structures}(b).
Serial episodes capture such a pattern
very well. Once again, we may need some additional time constraints. 
A useful constraint is that of inter-event time constraint.
In multi-neuron data, if we want to conclude that $A$ is causing $B$ to
fire, then $B$ cannot occur too soon after $A$ because there would be
some propagation delay and $B$ can not occur too much later than $A$
because the effect of firing of $A$ would not last indefinitely.
For example, we can prescribe that inter-event times should be in the range
of typical synaptic delay times.
Thus, serial episodes with proper inter-event time constraints can capture
ordered firing sequences which may be due to underlying functional connectivity.

Another important pattern in spiking data is that of synfire
chains \cite{Ikegaya2004}. This
consists of groups of synchronously firing neurons strung together with
tight temporal constraints, repeating often. The  structure shown
in Fig.~\ref{fig:episode-structures}(c) captures such a synfire chain. We can
think of this as a microcircuit where $A$ primes synchronous firing of $(B C D)$,
which, through $E$, causes synchronous firing of $(F G H I)$ and so on. When
such a pattern occurs often in the spike train data, parallel episodes like
$(B C D)$ and $(F G H I)$ become frequent. 
These parallel episodes, representing
synchrony, can be discovered with appropriate expiry time constraints. 
After discovering all such parallel episodes,  suppose we  replace all
recognized occurrences of each of these episodes by a new event in the data 
stream with a new symbol (representing the episode) for the event type and 
an appropriate time of occurrence.  Then if we 
discover serial episodes  on this new data stream, 
 we can unearth patterns such as 
synfire chains. We show later that our algorithms can discover such
synfire chains also efficiently.

In the language of partial orders, a serial episode corresponds to a chain or 
a totally ordered set. Similarly a parallel episode can be thought of as an 
anti-chain or that the underlying partial order is null. Synfire chains can be 
thought of as representing graded posets where all maximal chains have the same 
length. However, we do not formalize our episodes in this fashion because such a 
formalism is not particularly useful for our purposes. 

\begin{figure}
\centering
\epsfig{file=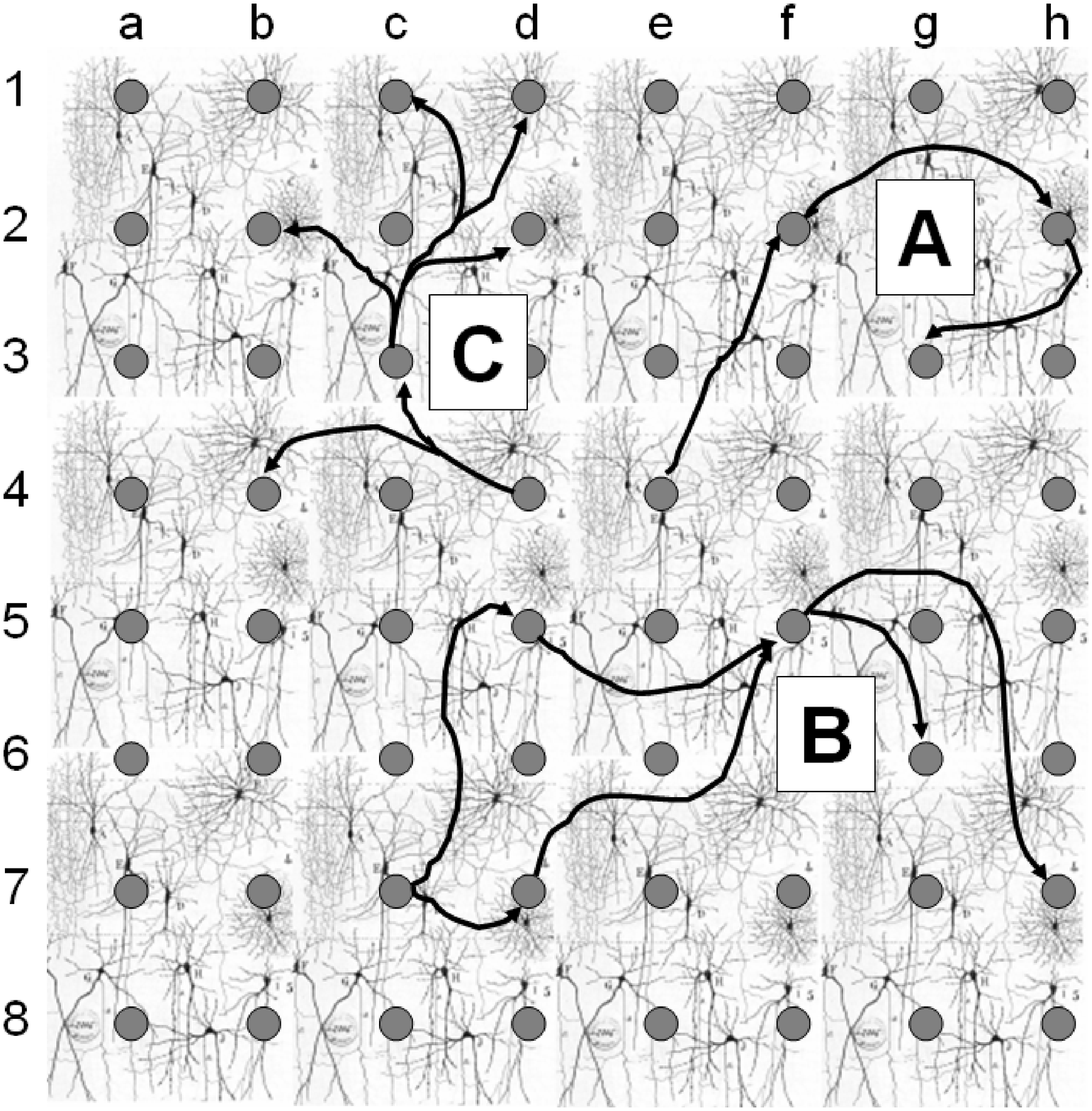, height=2.5in, width=2.5in}
\caption{Hypothetical Micro electrode array showing three networks that 
can manifest as the following episodes: 
[A]. $\mbox{e4}\rightarrow{\mbox{f2}}\rightarrow{\mbox{h2}}\rightarrow{\mbox{g3}}$, 
[B]. $\mbox{c7}\rightarrow{\mbox{(d5 d7)}}\rightarrow{\mbox{f5}}\rightarrow{\mbox{(g6 h7)}}$, 
[C]. $d4\rightarrow{\mbox{(b4 c3)}}\rightarrow{\mbox{(b2 d2 d1 c1)}}$.}
\label{fig:mea-episodes}
\end{figure}

Fig.~\ref{fig:mea-episodes} shows (in schematic fashion) how specific neuronal 
interconnections on a neural tissue underneath a micro electrode array can 
result in specific temporal patterns (in terms of our episodes) in the 
spike train data that is obtained from this micro electrode array recordings. 
In the figure, the electrodes are referred to by their position in the 2D grid 
and for three different neuronal connection patterns the resulting episode 
is shown. The connectivity pattern {\bf A} in the figure gives rise to a  
 serial episode. The other two connectivity patterns give rise to synfire 
chain type episodes which are a combination of parallel and serial episodes. 
Thus,  when we analyze data obtained through micro electrode array 
recordings using our algorithms, this is how the episodes are to be interpreted 
as connectivity patterns.

Summarizing the above discussion, we can say that discovering frequent 
serial and parallel episodes with appropriate temporal constraints 
would allow us to identify many of the important patterns in spike train data 
that capture underlying connectivity information. It is in this aspect that 
temporal data mining can play a vital role in analyzing spike train data.


\section{Discovering frequent episodes under temporal constraints}
\label{sec:algos}

In this section we describe our algorithms that discover frequent episodes 
under expiry and inter-event time constraints. Since algorithms for 
taking care of expiry time are available in case of serial episodes \cite{Srivats2006}, 
we consider the case of only parallel episodes under expiry time constraint. 
The inter-event time constraints are meaningful only for serial episodes and 
that is the case we consider. 
While describing 
the details of the algorithm we assume that the reader is familiar with the 
general structure of frequent episode discovery methods \cite{Mannila1997,Srivats2005}. 
However, for the sake of readers who are not familiar with these data mining 
techniques,  we also provide some intuitive explanation wherever possible.   

 A frequent episode is one whose frequency 
exceeds a user specified threshold. The overall objective is to find all 
frequent episodes. Counting of all possible episodes is infeasible in most 
real problems due to combinatorial 
explosion. As is common in such data mining methods, 
we use the same basic idea as in  level-wise Apriori-style \cite{Agrawal95} procedure 
which was first introduced in the context of frequent itemset mining. We briefly explain 
this idea below. 

Consider the problem of discovering all frequent serial episodes up to a given
size. The discovery process has two main steps. First, we build a set of candidate
episodes and next, we obtain the frequencies (i.e., count the non-overlapping
occurrences) of the candidates in the data stream 
 so that we can retain only those whose frequencies
are above the user-set threshold. The combinatorial explosion is in candidate generation 
because  the number of possible episodes of a given size increases exponentially with 
size. For now, let us
assume that there are no temporal constraints. The key observation is that an episode
can be frequent only if all its subepisodes are frequent. This is 
obvious because, for example, given two non-overlapping occurrences of
$A \rightarrow B \rightarrow C$, we have at least two non-overlapping
occurrences of each of its subepisodes. This immediately gives rise to a level-wise
procedure for discovering all frequent episodes. First we discover all frequent
1-node episodes. (This is simply a histogram of event types). Then we build a set
of candidate 2-node episodes such that the 1-node subepisodes of all candidates
are seen to be frequent. Now through one more pass over the data, we count the
non-overlapping occurrences of all the candidates and thus come out with frequent
2-node episodes. Now we combine only the frequent 2-node episodes to build
a candidate set of 3-node episodes and so on. Thus at stage $n$, using the
already discovered set of frequent $n$-node episodes, we build the set of
candidate $(n+1)$-node episodes and, by counting their occurrences in the data
(using one more pass over the data), we come out with frequent $(n+1)$-node
episodes. This procedure controls the combinatorial explosion because we are,
after all, interested only in episodes that occur sufficiently often. By choosing
a suitable frequency threshold, as the size of episodes grows, the number
of frequent episodes would come down. (It is highly unlikely that all large
random sequences occur often in the data). Because of this, the number of
candidates becomes  much less than the combinatorially possible number, as the
size of episodes grows. This is the basic idea of Apriori-style procedure used in 
most frequent pattern mining algorithms. 

When we impose temporal constraints, we need to ask whether such a procedure still works. 
It is easy to see that the property ( referred to as anti monotonicity 
property) of subepisodes being at least as frequent as the 
episode holds under expiry time constraint also. This is because every occurrence 
of the episode that completes within a given (expiry) time would contain occurrences 
of all the subepisodes which also complete within that time. However this property 
 does not hold under inter-event constraints. 
Suppose we need the time between any two consecutive events in any serial episode 
occurrence to be less than, say, $T_X$. We may have many occurrences of 
$A \rightarrow B \rightarrow C$ where
each pair of consecutive events occur within time  $T_X$, but there may
be no occurrence of the episode $A \rightarrow C$ such that the time between
the two events is less than $T_X$. However, as we shall see later on we can still use 
the level-wise procedure by restricting the check that subepisodes should be frequent,  
only to certain kind of subepisodes. The idea used here is similar to the concept 
of `loose anti-monotonicity' in the context of mining frequent itemsets under 
constraints \cite{Bonchi2007}.

Given that we can control the growth of candidates as the size of
episodes increases, the next question is how do we count the frequencies
of a set of candidate episodes. This is done by having a finite state
automaton for each episode such that it recognizes the occurrence of an
episode.\footnote{It may be noted here that the idea of finite state automata is 
essentially a conceptual tool for understanding the algorithms.}
 For example, for a serial episode, $A \rightarrow B \rightarrow C$, there 
would be an automaton that keeps waiting for an $A$ to transit into its first 
state and after that, waits for a $B$ to transit into its second state and so on. When 
this automaton transits into its final state, an occurrence would be complete and then 
we reset the automaton to wait for an $A$ again. For a parallel episode, conceptually, 
the states would be subsets of event types in the episode and the current state tells 
which all event types are yet to occur. We may have to keep track of more than one 
set of potential state transitions of the automaton and hence we may need multiple 
automata for each episode in the set of candidates. 
As we traverse the data, for each event (spike) we encounter, we make
appropriate state changes in all the automata (that are waiting for this event 
type) and whenever an automaton
transits to its end state we increment the count of the corresponding
episode. Thus, we can simultaneously count the occurrences of a set of candidates
using a single pass over the data. 
 Conceptually, all the algorithms for frequent episode discovery use
this general framework \cite{Mannila1997,Srivats2005,Srivats-kdd07}. 
We also use the same idea in our algorithms and also use essentially the 
same data structures as in \cite{Srivats2005} for keeping track of potential state
transitions of different automata. The main data structure is a {\em waits()} list 
which is indexed through event types. That is, {\em waits(A)} would store a list of 
(appropriate representations) of automata that are waiting for an occurrence of $A$. 
The number of active automata per episode that we
need (which is same as the temporary memory needed by the algorithm) depends on what all
types of occurrences we want to count. Restricting the count to only non-overlapped
occurrences makes the counting process also very efficient \cite{Srivats2005,Srivats-kdd07}.

The overall procedure for frequent episode discovery is given below as a 
pseudo code.  

\begin{algorithm}
\caption{\label{alg:Episode-discovery}Mining Frequent Episodes}
\begin{algorithmic}[1]
\STATE Generate an initial set of (1-node) candidate episodes (N=1)
\REPEAT
\STATE Count the number of occurrences of the set of (N-node) candidate 
episodes in one pass of the data sequence
\STATE Retain only those episodes whose count is greater than the
frequency threshold and declare them to be frequent episodes
\STATE Using the set of (N-node) frequent episodes, generate the next set
of (N+1-node) candidate episodes
\UNTIL{ There are no candidate episodes remaining}
\STATE Output all the frequent episodes discovered
\end{algorithmic}
\end{algorithm}

When we specify our algorithms in the next two subsections, we explain both candidate 
generation and frequency counting steps. 

\subsection{\label{sec:parallel-with-expiry}Parallel episodes with
expiry}

In this section we present an algorithm that counts the number of non-overlapped occurrences
of a set of parallel episodes in which all the constituting events occur within
time $T_{X}$ of each other. 
The algorithm here discovers parallel episodes with non-repeated event types. 
The pseudo-code for the algorithm is listed 
as \textit{Algorithm~\ref{alg:count-parallel-EXPIRY}} in the Appendix.

The algorithm takes as input, the set of candidate episodes, the event sequence, 
the expiry time, $T_{X}$,  
and the frequency threshold, and outputs the set of frequent episodes. An 
occurrence of a parallel episodes requires all its constituent nodes to appear 
in the event sequence in any order. At any given time, one needs to wait 
for all the nodes of the episode that remain to be seen. 
In the implementation of the algorithm here, instead of a single automaton 
waiting for a set of event types, we maintain separate entries for each distinct 
event type of the episode using a $waits(.)$ list indexed by event types. For 
event type $A$, each entry in the list  $waits(A)$ is of the form 
$(\alpha, count,init)$, where $\alpha$ is an episode waiting for an $A$, 
``$count$'' takes values 1 or 0 depending on whether an event of this  
type ($A$) has been seen or not, and $init$ indicates the latest time of 
occurrence of this event type.
The \emph{Algorithm}~\ref{alg:count-parallel-EXPIRY}, given as pseudo code, 
specifies the details of how the $waits$ list is updated as we traverse the data. 

An occurrence of an episode is complete when all required event types have been 
encountered at least once and all the 
event times (remembered by $init$ field) occur within $T_{X}$ of each 
other. An episode specific $counter$ is used to keep track of the event 
types already seen. When an occurrence is complete, 
the episode frequency count is incremented and 
all the entries (in the $waits(.)$ lists) for the episode are 
reinitialized.

If the expiry check fails, we cannot reject all the events types of the partial 
occurrence. This is because, in an occurrence of a parallel episode, the
constituent event types can occur in any order in the event stream. Only those
event types which have occurred before $(t_{i}-T_{X})$, should be rejected,
where $t_{i}$ is the time of the latest event type seen by the algorithm.
Effectively, any later occurrence of these events could possibly complete the
parallel episode (without violating the temporal constraint).

\subsubsection*{Candidate generation}

The candidate generation scheme is very similar to the one presented in 
\cite{Agrawal95} for itemsets. Let $\alpha$ and $\beta$ be two $k$-node frequent 
episodes having first $(k-1)$ nodes identical. The potential $(k+1)$-node 
candidate is generated by appending to $\alpha$ the $k^{th}$node of $\beta$. 
This new episode is declared as a $(k+1)$-node candidate if all its $k$-node 
subepisodes are already known to be frequent.

\subsection{\label{sec:interval-discovery-algo}Serial Episodes with Inter-event 
Constraints}

Under an inter-event time constraint, the time difference between 
  successive events in any occurrence 
have to be in a prescribed interval.  
The episode structure now consists of an ordered set of intervals besides the ordered set of 
event types. For example, a 4-node serial episode is 
now denoted as follows:
\begin{equation}
(A^{\underrightarrow{(t_{low}^{1},t_{high}^{1}]}}B^{\underrightarrow{(t_{low}^{2},t_{high}^{2}]}}C^{\underrightarrow{(t_{low}^{3},t_{high}^{3}]}}D)
\label{eq:episode-with-interval}
\end{equation}
In a given occurrence of episode $A\rightarrow$ $B\rightarrow$ $C\rightarrow$ $D$ 
let $t_{A}$, $t_{B}$, $t_{C}$ and $t_{D}$ denote the time of occurrence of corresponding 
event types. Then this is a valid occurrence of the serial episode with inter-event 
time constraint given by (\ref{eq:episode-with-interval}), 
if $t_{low}^{1}$ $<$ $(t_{B}-t_{A})$ $\le$ $t_{high}^{1}$,
$t_{low}^{2}$ $<$ $(t_{C}-t_{B})$ $\le$ $t_{high}^{2}$ and 
$t_{low}^{3}$ $<$ $(t_{D}-t_{C})$ $\le$ $t_{high}^{3}$.

In general, an $N$-node serial episode is associated with, $N-1$ inter-event 
constraints of the form $(t_{low}^{i},t_{high}^{i}]$. The algorithm we present 
is for generalized inter-event constraints. The user needs to specify only the 
granularity of search by providing a set of non-overlapped time intervals to serve 
as candidate inter-event time intervals. The algorithm 
 discovers all frequent episodes along with the choice of  
most appropriate inter-event intervals for each episode. 

\subsubsection{Candidate generation scheme}

As explained earlier, when we have inter-event time constraints, it is not 
necessary that all subepisodes have to be frequent for an episode to be 
frequent.  In 
the data sequence, if episode $(A^{\underrightarrow{(0,5]}}$  
$B^{\underrightarrow{(5,10]}}$ $C)$ is frequent, the sub-episodes 
$(A^{\underrightarrow{(0,5]}}$ $B)$ and $(B^{\underrightarrow{(5,10]}}$ $C)$ are 
also as frequent, but pairing event type $A$ with $C$ we would get 
$(A^{\underrightarrow{(?,?]}}$ $C)$ as a subepisode whose inter-event constraint 
is not intuitive.

\begin{figure}[!htb]
\centering
\input{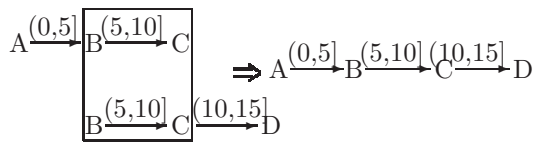}
\caption{\label{fig:Visualization-of-cand-gen-DISCOVERY}Visualization of
Candidate generation for serial episodes with inter-event constraints}
\end{figure}

The candidate episodes in this case are generated as follows. Let
$\alpha$ and $\beta$ be two $k$-node frequent episodes such that
by dropping the first node of $\alpha$ and the last node of $\beta$,
we get exactly the same $(k-1)$-node episode. 
 That is, the $(k-1)$-event
types match and also the $(k-2)$-intervals corresponding to the inter-event
constraints match. 
A candidate episode $\gamma$ is generated by copying
the $k$-event types and $(k-1)$-intervals of $\alpha$ into $\gamma$
and then copying the last event type of $\beta$ into the $(k+1)^{th}$
event type of $\gamma$ and the last interval of $\beta$ to the $k^{th}$
interval of $\gamma$. Fig.~\ref{fig:Visualization-of-cand-gen-DISCOVERY} shows 
the candidate generation process graphically.

It is easy to see that, under inter-event time constraints, an episode cannot be 
frequent unless such prefix and suffix subepisodes are frequent. It is this 
necessary condition that our candidate generation scheme exploits. 
This method of candidate generation is very similar to the ones used in case of mining of 
sequential patterns under similar inter-event constraints \cite{zaki-pap1,zaki-pap2}. 

\subsubsection{Counting episodes with generalized inter-event time constraint}

As already stated, the constraints are
in the form of intervals $(t_{low}^{i},t_{high}^{i}]$, in which the
inter-event times must lie. We first explain the difficulty in taking care 
of such constraints under the currently available algorithms for counting 
maximum number of non-overlapped occurrences of serial episodes. 

Suppose we want to 
count the non-overlapped occurrences of 
$A \rightarrow B \rightarrow C \rightarrow D$. In the data sequences there may be 
many instances of $A$, $B$, and $C$ before we get the first $D$ and the method of 
counting has to decide which of these to take while scanning the data from left to 
right for constituting an occurrence. There are two distinguishable approaches here 
which were used. Given a set of overlapped occurrences of the episode all ending 
on the same event, the one which starts earliest is called the left-most 
occurrence and the one that starts last is called the inner-most occurrence. 
(See \cite{Srivats2006} for more discussion). 
Consider the event sequence
\begin{equation}
\langle(A,1),(A,2),(B,4),(A,5),(C,10),(B,12),(C,13),(D,17)\rangle.
\label{eq:interval-discovery}
\end{equation}
This contains exactly one non-overlapped occurrence of the four node episode. 
Here, 
the left-most occurrence 
is $\langle(A,1),$ $(B,4),$ $(C,10),$ $(D,17)\rangle$ and the inner-most 
occurrence  is $\langle(A,5),$ $(B,12),$ $(C,13),$ 
$(D,17)\rangle$. Counting the left-most occurrence is efficient because we make 
a transition as soon one can be made and we do not need to remember any other 
possible transitions. On the other hand, for counting the inner-most occurrence 
we have to remember multiple transition possibilities. However, for implementing 
expiry time constraint, we have to count the inner-most occurrence because the 
left-most occurrence may not satisfy the expiry time constraint. (If 
any of the occurrences in this set of occurrences satisfy the expiry time 
constraint then the inner-most would). There are  algorithms for 
counting non-overlapped occurrences of serial episodes in both these fashions. 
When we have inter-event constraints, we can not get maximum number of non-overlapped 
occurrences by counting only either left-most or inner-most. Suppose, in this 
example, the inter-event constraints are given as  
 $(A^{\underrightarrow{(0,5]}}$ $B^{\underrightarrow{(5,10]}}$ $C^{\underrightarrow{(0,5]}}$ $D)$.
In the given event sequence, 
 only the occurrence $\langle(A,2),$ $(B,4),$ $(C,13),$ 
$(D,17)\rangle$  satisfies the inter-event interval constraints. 
This is the reason we need to modify the counting scheme so as to  
 remember more number of potential transitions than needed to count 
inner-most occurrences. 

The counting algorithm is listed as Algorithm 2 in the Appendix. 
The  algorithm presented uses $waits$ lists indexed by event types 
  as the basic data-structure.  
The entries in the $waits$ lists are structures called $node$s. For each episode we have a 
doubly linked list of $node$ structures with a $node$ corresponding to each of the event 
types and arranged in the same order as that of the episode. 
The $node$ structure has a $tlist$ field that stores the times of 
occurrence of the event-type represented by its corresponding $node$. 
For example, in the event sequence given by (\ref{eq:interval-discovery}), 
the $node$ representing $A$, after $t=5$, would have $tlist=\{(A,1),(A,2),(A,5)\}$. 
Other field in the $node$ structure is $visited$, 
which is a boolean field that indicates whether the event type is seen at least once.

\begin{figure}[!htb]
\centering
\epsfig{file=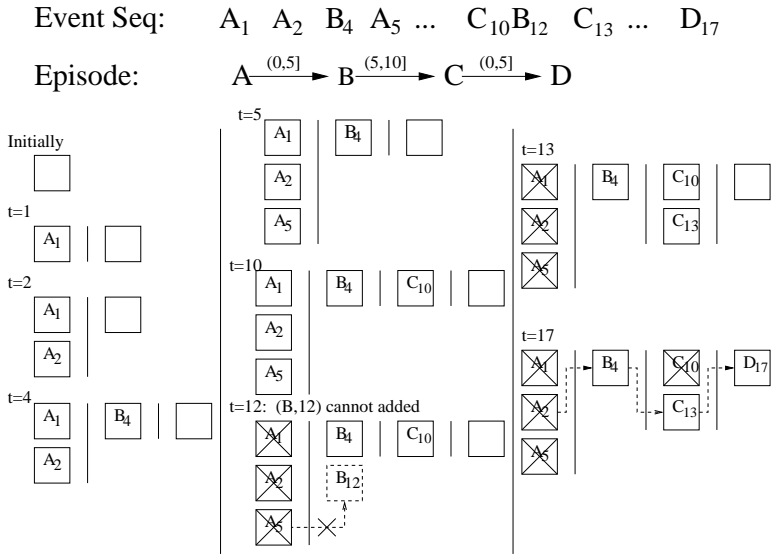, width=4in}
\caption{\label{fig:Visualization-of-interval-DISCOVERY}Visualization of
Algorithm~\ref{alg:count-serial-INTERVAL-discovery}}
\end{figure}

On seeing an event type $E_{i}$, the algorithm iterates over list $waits(E_{i})$ 
and updates each $node$ in the list. We explain the procedure for updating the $node$s 
by considering the the example sequence given in 
(\ref{eq:interval-discovery}) and the episode 
$\alpha=(A^{\underrightarrow{(0,5]}}$ $B^{\underrightarrow{(5,10]}}$ 
$C^{\underrightarrow{(0,5]}}$ $D)$. Working of the algorithm in this example is 
illustrated in Fig.~\ref{fig:Visualization-of-interval-DISCOVERY}.

The $waits$ lists are initialized by adding the $node$s corresponding
to first event type of each episode in the set of candidates to the 
corresponding $waits(.)$ list. In the example, let the $node$
tracking event type $A$ be denoted by $node_{A}$, and so on. 
Initially $waits(A)$ contains $node_{A}$.
The boxes in 
Fig.~\ref{fig:Visualization-of-interval-DISCOVERY}
represent an entry in the $tlist$ of a $node$. An empty box is one
that is waiting for the first occurrence of an event type. On seeing
$(A,1)$, it is added to $tlist$ of $node_{A}$, and $node_{B}$
is added to $waits(B)$. At any time, \emph{the $node$ structures
are waiting for all event types that have been already seen and the
next unseen event type}.

The algorithm is now waiting for an occurrence of a $B$ and an $A$
as well. At $t=4$, the first occurrence of a $B$ is seen. The $tlist$
of $node_{A}$ is traversed to find at least one occurrence \emph{of
$A$,} such that $t_{B}-t_{A}\in(0,5]$. Both $(A,1)$ and $(A,2)$
satisfy the inter-event constraint and hence, $(B,4)$ is accepted
into the $node_{B}.tlist$. The rule for accepting an occurrence of
an event type (which is not the first event type of the episode) is
that \emph{there must be at least one occurrence of the previous event
type} (in this example $A$) \emph{which can be paired with the occurrence
of the current event type} (in this example $B$) \emph{without violating
the inter-event constraint}.  
After seeing the first occurrence of $B$, $node_{C}$ is added to $waits(C)$.
Using the above rules the algorithms accepts $(A,5)$, $(C,10)$ into
the corresponding $tlist$s. At $t=12$, for $(B,12)$ none of the
entries in $node_{A}.tlist$ satisfy the inter-event constraint for
the pair $A\rightarrow B$. Hence $(B,12)$ is not added to the $tlist$
of $node_{B}$. Rest of the steps of the algorithm are illustrated in the figure.

If an occurrence of event type is added to $node.tlist$, it is because there 
exist events for each event type from the first to the event type corresponding 
to the $node$, which satisfy the respective inter-event time constraints.
An occurrence of episode is complete when an occurrence of the last
event type can be added to the $tlist$ of the last $node$ structure
of the episode.

The $tlist$ entries shown crossed out in the figure are the ones
that can be deallocated from the memory. 
In the example, at $t=12$, when the algorithm
tries to insert $(B,12)$ into $node_{B}.tlist$, the list of $tlist$
entries for occurrences of $A$ is traversed. $(A,1)$ with inter-event
constraint $(0,5]$ can no longer be paired with a $B$ since the
inter-event time duration for any incoming event exceeds $5$, hence
$(A,1)$ can be safely removed from the $node_{A}.tlist$. This holds
for $(A,2)$ and $(A,5)$ as well. In this way the algorithm frees
memory wherever possible without additional processing burden.

In order to track episode occurrences we need to store sufficient back references 
in data structures to back track each occurrence. 
This adds some memory overhead, but tracking may be useful in visualizing 
the discovered episodes.

\section{Simulation Results}
\label{sec:res}

In this section we present some results obtained with our algorithms for analyzing
spike-train data. We mainly discuss results obtained on synthetic data generated through
a simulation model. We also present some results on data gathered from experiments
on neural cultures. The main
reason for using  simulator-generated data is that here we can have control
on the kind of patterns that the data contains and can thus check whether our
algorithms discover the `true' patterns. Also, by generating many sets of random 
data we can study the statistical properties of the algorithms. 
The simulation model is intended to
generate fairly realistic spike trains. For this we simulate a network
of neurons where each neuron is modeled as an inhomogeneous 
 Poisson process whose rate changes
with the input received by the neuron. 
On simulator-generated data we present results to show
that our algorithms can discover different types of embedded patterns. We also present
some empirical results to argue that the patterns discovered would be statistically
significant.

\subsection{The spike data generation model}

For the data generation, we use a simulator where each neuron is modeled as an
inhomogeneous poisson process.
The neurons are interconnected and there is a weight
attached to each interconnection or synapse. The rate of firing of
each neuron is time-varying  and it is dependent on the weighted sum of spikes 
received (over a time window) as input  by the 
neuron from the others through the synapses. 
The rate is updated every $\Delta T$ time units. 
We use two different functions to change the firing rate of a neuron in response to 
inputs received and test our methods in both scenarios. 

As explained earlier, the intuitive reason why temporal data mining may be 
effective for spike data analysis  is the following. 
Suppose neuron $A$ is connected to $B$ and the neuron $B$ to $C$ through excitatory synapses 
each of synaptic delay of about $T$. When $A$ fires, because of the excitatory synapse, 
the firing rate of $B$ would go up after a delay of about $T$ and similarly for the 
$B$ to $C$ connection. Then, irrespective of the 
underlying mechanism of changing the rate as a function of input, we would essentially 
have high conditional probability of seeing a spike from $B$ 
at time $T$ given $A$ has 
spiked at time zero and similarly for the $B$ to $C$ connection. 
This would make an occurrence of the serial episode, 
$A\rightarrow B \rightarrow C$ with inter-event constraint of about $T$, much more likely 
than that for any other triplet of non-interconnected neurons. Thus if a 3-node episode like 
this has high enough frequency (in terms of number of non-overlapped occurrences) then 
it is very likely that it indicates such a connectivity pattern. As long as the underlying 
model of interconnected neurons results in a high value of the conditional probability 
referred to above, the data mining technique would be able to discover the connectivity 
pattern irrespective of the specific underlying mechanism of interaction 
among neurons.  (Note that if we 
postulate a specific stochastic model for the interacting neurons and then estimate the 
model parameters, the analysis results are tightly coupled with the 
assumed model). It is simply to 
illustrate this that we have used two different ways of updating the rate of firing 
based on the input. The specific mechanisms may or may not be biologically plausible; but  
are used here to demonstrate model independence of the proposed technique. 

The two methods of updating the firing rate are as follows. The first one uses a sigmoidal 
relationship. Specifically we use 
\begin{equation}
\lambda_j(k) = \frac{\lambda_{1j}}{1 + \exp{(-I_j(k) + d)}}
\label{eq:lambda-update}
\end{equation}
where $\lambda_j(k)$ is the firing rate of $j^{th}$ neuron at time $k \Delta T$ 
and $I_j(k)$ is its total input at that time. $I_j(k) = \sum_i O_i(k) w_{ij}$ 
where $O_i(k)$ is the output of $i^{th}$ neuron  (as seen by 
the $j^{th}$ neuron) at time $k \Delta T$ 
 and $w_{ij}$ is the weight of synapse from $i^{th}$ to $j^{th}$ neuron.
$O_i(k)$ is taken to be the number of spikes by the $i^{th}$ neuron in the time  
interval $(\;(k-h-1) \Delta T, \ (k-h) \Delta T]$ where $h$ represents the 
synaptic delay in units of $\Delta T$.   
The offset parameter $d$ determines the resting spiking 
rate $\lambda_{0j}$ (i.e. the firing rate when input is zero). 
This is the quiescent firing rate 
(or the noise level) in the system. We choose $\lambda_{1j}$ to ensure that we 
reach the desired high firing rate on receiving the amount of input that should prime this 
neuron to fire. An absolute refractory period is also implemented. 
This is the short time after a spike in which the neuron cannot respond to 
another stimulus.

In the second method of updating,  the firing rate is low if 
input is low and increases linearly with input till it reaches the desired firing 
rate for full input and saturates at that rate. Specifically we use
\begin{eqnarray}
\lambda_j(k) &=& 0 \ \ \mbox{~if~} \ I_j(k) \leq - \frac{\lambda_{0j}}{a} \nonumber \\
 &=& a I_j(k) + \lambda_{0j} \ \ \mbox{~if~} \ 
- \frac{\lambda_{0j}}{a} \leq I_j(k) \leq I_{1j} \nonumber \\
 &=& \lambda_{1j} \ \ \mbox{~if~} \ I_j(k) > I_{1j}
\label{eq:lambda-update2} 
\end{eqnarray}
Here, $a$ is a suitable constant that determines slope of the linear region, $\lambda_{0j}$ 
represents the quiescent firing rate of neuron $j$ which is the firing rate when input is 
zero and $\lambda_{1j}$ represents the high firing rate which is reached on full input 
(which is represented by the parameter, $I_{1j}$). 
The input to the $j^{th}$ neuron at time 
$k \Delta T$, $I_j(k)$, is determined as in the earlier case.  

To employ either model, in the simulator we  specify the quiescent rate of firing 
 (denoted by $\lambda_0$)  and the required conditional probability of the 
receiving neuron firing (denoted by $\rho$) when it 
receives the full expected input. 
The quiescent firing rate determines $d$ in the 
first model and $\lambda_{0j}$ in the second model. The conditional probability 
specified determines the high end of firing rate we want to reach. Now we choose 
the weight values for the intended connections so that on receiving the full input 
we reach the high firing rate needed to satisfy the specified conditional probability. 
 
We use these models to generate data with different patterns as follows.
Let N denote the total number of neurons in the
system. (We have generated data with N=26, 64 and 100).
First we randomly interconnect the neurons. For each neuron
we randomly choose $M$ other neurons to interconnect it with. (We used
$M= N/2$ and $N/4$). The weight attached to each synapse is set randomly using a
uniform distribution over $[-c, \; c]$.\footnote{Using   
 either model, given a weight of interconnection we can calculate 
the conditional probability of the receiving neuron firing in a 
given interval of time conditioned on spiking 
by the sending neuron. Hence, we can always specify weight of interconnection in 
terms of the conditional probability it represents. In some simulations later on, we 
specify in this fashion the random weights used  because it is more intuitive}.  
When we want to embed any specific pattern, then, we set the weights of the required
connections between neurons to a higher value. The weight value depends on the type of 
pattern to be embedded. 
The patterns we want to embed are the kind shown in Fig.~\ref{fig:episode-structures}.
These are realizable by essentially two types of pattern dependent
interconnections between neurons. One
is where a neuron primes one (as in a serial episode) or many (as in a parallel
episode) neurons. Here a high conditional probability ($\rho$) 
is assigned to each interconnection
that forms the part of a pattern and the corresponding weight is 
determined by requiring that one spike (in the
appropriate interval) by the priming neuron would increase the firing rate of the
receiving neuron  so that in the next $\Delta T$ interval (after the synaptic delay time) the
receiving neuron spikes at least once with probability $\rho$. (The required firing 
rate is determined based on $\rho$). The other kind of
interconnection is where many neurons together prime one neuron (which is used
in Synfire chains). Here, the weight of each connection is set in such a way that
only if each of the input neurons spikes once in the appropriate interval then the
firing rate of the receiving neurons would go up to the high value. (If only a few of
the input neurons fire, then the firing rate of the receiving neuron goes up but
not all the way up to the high level).

The simulator runs as follows. 
At any given time, we have some spiking rates for all the neurons. Using this,
spikes are generated (as a Poisson process) for the next $\Delta T$ time interval 
for every neuron. (Since we are using a Poisson process model, the time of a
spike would be a real number). Then, using the spiking history, we calculate the
inputs into each neuron and update the spiking rate for each neuron using
eq.(\ref{eq:lambda-update}) or (\ref{eq:lambda-update2}) depending on which model 
we are using. Then with the new spiking rates we generate spikes from
every neuron for the next $\Delta T$ and so on. We also implement an absolute refractory
period for each neuron. Once the spikes are generated for a $\Delta T$ interval, starting
with the second spike we remove spikes one by one if they are closer than $\tau_r$ to the
previous spike where $\tau_r$ is the refractory time.

The network contains both random interconnections among neurons 
as well as interconnections (with larger weights) that together constitute some patterns 
embedded. 
The weights of random connections are set using a mean-zero distribution and hence,
in an expected sense all neurons keep firing at the `noise' rate of $\lambda_0$. However,
since the actual input can still assume small positive and negative values, this
background firing rate would also be fluctuating around $\lambda_0$. Since all firings
are stochastic, even when a pattern is embedded, the entire patterned firing sequence
will not always happen. Also, within a pattern of firing of neurons (as per the embedded
pattern), there would be other neurons that would be spiking randomly. Though we calculate 
the input into neurons and their firing rates at only discrete time points (separated 
by $\Delta T$), since each neuron's spikes are (inhomogeneous) Poisson, the actual 
times of spikes would be continuous.  That is, the time of spikes would not be integral 
multiples of $\Delta T$. ( We note in passing that, due to
our implementing of refractory period, the actual firings of neurons are not 
strictly Poisson).

For the simulations discussed here, we generally use the following values for parameters:
 $\lambda_0 = 20 Hz$, $\rho = 0.95$. We have chosen
$\Delta T = 1$ ms and chosen the refractory period ($\tau_r$) also the same.
(This would mean that in any $\Delta T$ interval there would be at most one spike from any
neuron). For calculation of input into a neuron, we mostly use $h=5$ 
which implies a synaptic delay of 5 ms. In some examples, where we want to 
choose different synaptic delays for different connections we change the value of 
$h$ as needed. Similarly we can use different $\lambda_0$ for different neurons and 
different $\rho$ for different patterned interconnections when needed.

\subsection{Discovering Network patterns}

In this section, we demonstrate how we can obtain useful information
about the structure of the underlying network using combination of
serial and parallel episode discovery. Using the simulation model
described above, we can embed different types of network patterns.
Fig.\ref{fig:episode-structures} shows examples of types of inter connections 
we make to embed different patterns. For this, in the simulator 
we make these required connections between neurons have high weights as 
explained in the previous subsection. 
(In addition there are also random interconnections among neurons).
 
We discuss three examples to illustrate that our method 
is very effective in unearthing the underlying connectivity pattern.
For each example we use a specific connectivity pattern and  generate 
spike data using our simulator with these embedded patterns. We then use our algorithms 
to discover frequent parallel and serial episodes with different 
temporal constraints. As explained earlier, synchronous firing patterns are 
well captured by parallel episodes with appropriate expiry time constraint, ordered 
firing patterns are captured by serial episodes with inter-event constraints and 
synfire chains are captured by a combination of parallel and serial episodes. We 
illustrate all this in the examples. 

\subsubsection*{Example 1}
\begin{figure}[!htb]
\centering
\input{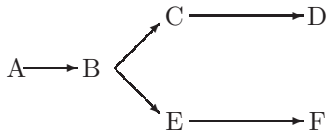}
\caption{\label{fig:pattern-1}Network pattern for Example 1}
\end{figure}

In a 26 neurons network (where each neuron corresponds to an alphabet)
we embed the pattern shown in Fig.\ref{fig:pattern-1}. The simulation
is run for 50 seconds. (In this time approximately 25,000 spikes are generated 
when we use eq.~(\ref{eq:lambda-update}) and about 20,000 spikes are generated 
when we use eq.~(\ref{eq:lambda-update2}) ). The
synaptic delay is set to  5 ms. We have chosen 
$\Delta T = 1$ ms and have taken refractory time also to be 
the same.

\begin{table}[!htb]
\centering
\begin{tabular}{|r|r|r|r|l|}
 \hline
Episode & Freq. & Time & Size & Patterns \\
expiry & Th. & (sec) & (No.) & Discovered \\
\hline
0.0001 & 0.01 & 0.23 & 1(26) & \small{no episode of 2} \\
 & & & & \small{or more nodes} \\ \hline
0.001 & 0.01 & 0.29 & 2(2) & \small{E C : 799; F D : 624} \\ \hline
0.002 & 0.01 & 0.28 & 2(2) & \small{E C : 804; F D : 643} \\ \hline
0.007 & 0.01 & 0.37 & 4(1) & \small{F E D C : 615} \\ \hline
\end{tabular}  
\caption{\label{tab:pattern-1-parallel}Parallel episodes for Example 1. The 
data is generated using (\ref{eq:lambda-update}) for updating firing rate of 
neurons}
\end{table}

\begin{table}[!htb]
\centering
\begin{tabular}{|r|r|r|r|l|}
 \hline
Episode & Freq. & Time & Size & Patterns \\
expiry & Th. & (sec) & (No.) & Discovered \\
\hline
0.0001 & 0.01 & 0.23 & 1(26) & \small{no episode of 2} \\
 & & & & \small{or more nodes} \\ \hline
0.001 & 0.01 & 0.25 & 2(2) & \small{E C : 652; F D : 502} \\ \hline
0.002 & 0.01 & 0.17 & 2(2) & \small{E C : 683; F D : 603} \\ \hline
0.007 & 0.01 & 0.48 & 4(1) & \small{F E D C : 525} \\ \hline
\end{tabular}
\caption{\label{tab:pattern-1-parallel-2}Parallel episodes for Example 1. The data 
is generated using (\ref{eq:lambda-update2}) for updating firing rate of neurons.}
\end{table}

\begin{table}[!htb]
\centering
\begin{tabular}{|r|r|r|r|l|}
 \hline
Inter-event & Freq. & Time & Size & Patterns \\
interval & Th. & (sec) & (No.) & Discovered \\
 \hline
0.000-0.001 & 0.01 & 0.29 & 2(4) & \small{C E : 410; E C : 400} \\
 &  &  &  & \small{D F : 329; F D : 303} \\ \hline
0.000-0.002 & 0.01 & 0.31 & 2(4) & \small{C E : 422; E C : 408} \\
 &  &  &  & \small{D F : 348; F D : 323} \\ \hline
0.002-0.004 & 0.01 & 0.26 & 1(26) & \small{no 2 or more} \\ 
	&	&	&	& \small{node episodes} \\ \hline
0.004-0.006 & 0.01 & 0.29 & 4(4) & \small{A B C D : 597} \\
 &  &  &  & \small{A B E F : 589} \\
 &  &  &  & \small{A B E D : 530} \\
 &  &  &  & \small{A B C F : 530} \\ \hline
\end{tabular}  
\caption{Serial episodes for Example 1. The 
data is generated using (\ref{eq:lambda-update}) for updating firing rate of 
neurons}
\label{tab:pattern-1-serial}
\end{table}

\begin{table}[!htb]
\centering
\begin{tabular}{|r|r|r|r|l|}
 \hline
Inter-event & Freq. & Time & Size & Patterns \\
interval & Th. & (sec) & (No.) & Discovered \\
 \hline
0.000-0.001 & 0.01 & 0.27 & 2(4) & \small{C E : 350; E C : 321} \\
 &  &  &  & \small{D F : 254; F D : 263} \\ \hline
0.000-0.002 & 0.01 & 0.31 & 2(4) & \small{C E : 382; E C : 365} \\
 &  &  &  & \small{D F : 330; F D : 338} \\ \hline
0.002-0.004 & 0.01 & 0.26 & 1(26) & \small{no 2 or more} \\
        &       &       &       & \small{node episodes} \\ \hline
0.004-0.006 & 0.01 & 0.29 & 4(4) & \small{A B C D : 259} \\
 &  &  &  & \small{A B E F : 248} \\
 &  &  &  & \small{A B E D : 220} \\
 &  &  &  & \small{A B C F : 213} \\ \hline
\end{tabular}
\caption{Serial episodes for Example 1. The 
data is generated using (\ref{eq:lambda-update2}) for updating firing rate of 
neurons}
\label{tab:pattern-1-serial-2}
\end{table}

The sequence is then mined for frequent parallel episodes with
different expiry times. The results are given in Tables
\ref{tab:pattern-1-parallel} and \ref{tab:pattern-1-parallel-2} for the two cases 
of data generated with models (\ref{eq:lambda-update}) and  (\ref{eq:lambda-update2}). 
The tables show the expiry time used, 
the frequency threshold, time taken by the algorithm on a Intel dual core 
PC running at 1.6 GHz, the size of the largest frequent episode discovered and 
the number of episodes of this size along with the actual 
episodes and their frequencies. (In the 
tables, all times are shown in seconds). We follow 
the same structure for all the tables. The frequency threshold is expressed as
a fraction of the entire data length. A threshold of 0.01 over a data
length of 25,000 spike events requires an episode to occur at least 250
times before it is declared as frequent. In the algorithm, this frequency 
threshold is what is used for episodes of size 1. As the size of episodes 
increases, the expected frequency decreases and hence we should suitably 
decrease the threshold. (See \cite{Srivats2005} for a more precise 
account of this). Hence, we reduce the frequency threshold by a factor of 0.9 
for each successively larger size of episodes. From Table
\ref{tab:pattern-1-parallel} it can be seen that $(CE)$ and $(DF)$
turn out to be the only frequent parallel episodes if the 
expiry time is 1 to 2 ms. If the expiry time is too small (less than 1 ms), we get 
no frequent episodes. On the other hand, if we increase 
the expiry time to be 7 ms which is greater than a synaptic delay, 
then even $(FEDC)$ turns out to be a parallel episode. From 
Table~\ref{tab:pattern-1-parallel-2}, it can be seen that the results obtained 
are the same even when we use the linear model given by 
(\ref{eq:lambda-update2}) for changing the firing rates of neurons. The only 
difference is some minor variation in  the actual number of non-overlapped 
occurrences. As long as the frequency is high enough the actual number is 
unimportant in discovering frequent episodes. These results  show that by using 
appropriate expiry time, parallel episodes discovered 
capture connectivity structure representing synchronous firing patterns.

The results of serial episode discovery are shown in Tables 
\ref{tab:pattern-1-serial} and \ref{tab:pattern-1-serial-2} for the two models of 
updating firing rates of neurons. With an inter-event constraint of 4-6 ms, 
 we discover all paths in the network (Fig.
\ref{fig:pattern-1}). When we prescribe that inter-event time be less 
than 2 ms (while synaptic delay is 5 ms), we get nodes in the 
same level as our serial episodes. If we use intervals of 2-4 ms, we 
get no episodes because synchronous firings mostly occur much closer 
and firings related by a synapse have a delay of 5 ms. Thus, using 
inter-event time constraints, we can get fair amount of information of the 
underlying connection structure that contributes to ordered firings.  
It may seem surprising that we also discover
$A\rightarrow B$ $\rightarrow C$ $\rightarrow F$ and $A\rightarrow B$
$\rightarrow E$ $\rightarrow D$ when we use 4--6 ms constraint. 
This is because, the network structure is 
such that $D$ and $F$ fire about one synaptic delay time 
after the firing of $C$ and $E$. Thus, the serial episodes give the sequential 
structure in the firings which could, of course, be generated by different 
interconnections. The frequent episodes discovered 
provide a handle to unearthing the hierarchy seen in the data (i.e.
which events co-occur and which ones follow one another). As seen from the 
tables, we get similar results with both methods of updating the firing rate. 

\subsubsection*{Example 2}
In this example we consider the network connectivity pattern as shown in
Fig.~\ref{fig:episode-structures}(c). As stated earlier, this is an example 
of possible network connectivity that can generate patterns known as 
Synfire chains. We use the same parameters in the simulator as in Example 1
 and generate spike trains data using this connectivity pattern.  As earlier, 
we generate data sets using the two different models for 
updating firing rates. 
Tables~\ref{tab:pattern-3-parallel} and \ref{tab:pattern-3-parallel-2}  show 
 the parallel episodes discovered  
 and Tables~\ref{tab:pattern-3-serial} and \ref{tab:pattern-3-serial-2} show 
 the serial episodes discovered under the two methods of updating firing rates. 
 From the tables, it is easily seen 
that parallel episodes with expiry time of 1 ms and 
serial episodes with inter-event time constraint of about one synaptic delay, 
together give good information about underlying network structure. Since the 
synaptic delay is taken to be 5ms, when we use an inter-event time constraint of 
4--6ms, all paths in the network become frequent serial episodes. There are 24 such 
serial episodes and hence the tables list only a few. Once again, 
the results under the two models of updating firing rates, are identical. 

In this example, 
we illustrate how our algorithms can discover synfire chain patterns. We first 
discover all parallel episodes with expiry time 1 ms. Then for  
each frequent parallel episode,  
we replace each of its occurrences in the data stream by a new event with event type 
being the name of the parallel episode. This new event is put in with a time 
of occurrence which is the mean time in the episode occurrence. We then discover 
all serial episodes with different inter-event time constraints. The 
results obtained with this method are shown in Tables~\ref{tab:pattern-3-synfire} 
and \ref{tab:pattern-3-synfire-2}. 
As can be seen, 
the only pattern we discover is the underlying synfire chain. (In the tables, 
the size of this is shown as the size of the serial episode where some 
of the nodes may be parallel episodes). This example shows 
that by proper combination of parallel and serial episodes, we can obtain fairly 
rich pattern structures which are of interest in neuronal spike train analysis. 

\begin{table}[!htb]
\centering
\begin{tabular}{|r|r|r|r|l|}
 \hline
Episode & Freq. & Time & Size & Patterns \\
expiry & Th. & (sec) & (No.) & Discovered \\
\hline
0.001 & 0.01 & 0.15 & 4(1) & \small{L K : 307} \\
 &  &  &  & \small{C B D : 293} \\
 &  &  &  & \small{H G F I : 268} \\
 &  &  &  & \small{rest are} \\
 &  &  &  & \small{sub-episodes} \\ \hline
\end{tabular}  
\caption{\label{tab:pattern-3-parallel}Parallel episodes for Example 2. 
The data is generated using (\ref{eq:lambda-update}) for updating firing rate of 
neurons}
\end{table}

\begin{table}[!htb]
\centering
\begin{tabular}{|r|r|r|r|l|}
 \hline
Episode & Freq. & Time & Size & Patterns \\
expiry & Th. & (sec) & (No.) & Discovered \\
\hline
0.001 & 0.01 & 0.1 & 4(1) & \small{L K : 228} \\
 &  &  &  & \small{C B D : 204} \\
 &  &  &  & \small{H G F I : 195} \\
 &  &  &  & \small{rest are} \\
 &  &  &  & \small{sub-episodes} \\ \hline
\end{tabular}
\caption{\label{tab:pattern-3-parallel-2}Parallel episodes for Example 2. The data 
is generated using (\ref{eq:lambda-update2}) for updating firing rate of neurons.}
\end{table}

\begin{table}[!htb]
\centering
\begin{tabular}{|r|r|r|r|l|}
 \hline
Inter-event & Freq. & Time & Size & Patterns \\
interval & Th. & (sec) & (No.) & Discovered \\
\hline
0.002-0.004 & 0.01 & 0.16 & 1(26) & \small{no episodes of 2} \\
 & & & & \small{or more nodes} \\ \hline
0.004-0.006 & 0.01 & 0.47 & 6(24) & \small{A D E H J K : 195} \\
 &  &  &  & \small{A D E I J K : 194} \\
 &  &  &  & \small{A D E H J L : 193} \\
 &  &  &  & \small{A C E H J K : 192} \\ \hline
0.006-0.008 & 0.01 & 0.16 & 1(26) & no episodes of 2 \\
 & & & & or more nodes \\ \hline
\end{tabular}  
\caption{Serial episodes for Example 2. The 
data is generated using (\ref{eq:lambda-update}) for updating firing rate of 
neurons}
\label{tab:pattern-3-serial}
\end{table}

\begin{table}[!htb]
\centering
\begin{tabular}{|r|r|r|r|l|}
 \hline
Inter-event & Freq. & Time & Size & Patterns \\
interval & Th. & (sec) & (No.) & Discovered \\
\hline
0.002-0.004 & 0.01 & 0.19 & 1(26) & \small{no episodes of 2} \\
 & & & & \small{or more nodes} \\ \hline
0.004-0.006 & 0.01 & 0.35 & 6(24) & \small{A D E H J K : 145} \\
 &  &  &  & \small{A D E I J K : 134} \\
 &  &  &  & \small{A D E H J L : 132} \\
 &  &  &  & \small{A C E H J K : 121} \\ \hline
0.006-0.008 & 0.01 & 0.1 & 1(26) & no episodes of 2 \\
 & & & & or more nodes \\ \hline
\end{tabular}
\caption{Serial episodes for Example 2. The 
data is generated using (\ref{eq:lambda-update2}) for updating firing rate of 
neurons}
\label{tab:pattern-3-serial-2}
\end{table}

\begin{table}[!htb]
\centering
\begin{tabular}{|r|r|r|r|l|}
 \hline
Inter-event & Freq. & Time & Size & Patterns \\
interval & Th. & (sec) & (No.) & Discovered \\
\hline
0.002-0.004 & 0.01 & 0.11 & 1(20) & \small{no episodes of} \\
 & & & & \small{2 or more nodes} \\ \hline
0.004-0.006 & 0.01 & 0.14 & 6(1) & \small{A [C B D] E} \\
 & & & & \small{[H G F I] J [L K] : 137} \\ \hline
0.006-0.008 & 0.01 & 0.12 & 1(20) & \small{no episodes of} \\
 & & & & \small{2 or more nodes} \\ \hline
\end{tabular}  
\caption{Synfire episodes for Example 2. The 
data is generated using (\ref{eq:lambda-update}) for updating firing rate of 
neurons}
\label{tab:pattern-3-synfire}
\end{table}

\begin{table}[!htb]
\centering
\begin{tabular}{|r|r|r|r|l|}
 \hline
Inter-event & Freq. & Time & Size & Patterns \\
interval & Th. & (sec) & (No.) & Discovered \\
\hline
0.002-0.004 & 0.01 & 0.2 & 1(20) & \small{no episodes of} \\
 & & & & \small{2 or more nodes} \\ \hline
0.004-0.006 & 0.01 & 0.3 & 6(1) & \small{A [C B D] E} \\
 & & & & \small{[H G F I] J [L K] : 108} \\ \hline
0.006-0.008 & 0.01 & 0.1 & 1(20) & \small{no episodes of} \\
 & & & & \small{2 or more nodes} \\ \hline
\end{tabular}
\caption{Synfire episodes for Example 2. The 
data is generated using (\ref{eq:lambda-update2}) for updating firing rate of 
neurons}
\label{tab:pattern-3-synfire-2}
\end{table}

\subsubsection*{Example 3}
\begin{figure}[!htb]
\centering
\input{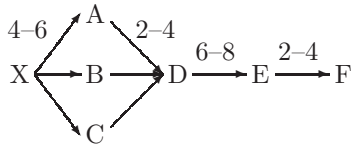}
\caption{\label{fig:pattern-2}Network Pattern for Example 3}
\end{figure}

In this example, we choose a network pattern where different pairs of 
interconnected neurons can have different synaptic delays 
and we demonstrate the ability of our algorithm to 
automatically choose appropriate inter-event intervals. 
The pattern is shown in Fig.
\ref{fig:pattern-2}, where we have different synaptic delays 
as indicated on the figure, for different inter-connections.

\begin{table}[!htb]
\centering
\begin{tabular}{|r|r|r|r|l|}
 \hline
Episode & Freq. & Time & Size & Patterns \\
expiry & Th. & (sec) & (No.) & Discovered \\
\hline
0.001 & 0.01 & 0.28 & 3(1) & \small{A B C : 614} \\ \hline
0.002 & 0.01 & 0.25 & 3(1) & \small{A B C : 617} \\ \hline
0.004 & 0.01 & 0.28 & 4(1) & \small{A B C D : 537} \\ \hline
0.006 & 0.01 & 0.32 & 4(2) & \small{X A B C : 602} \\
 &  &  &  & A B C D : 542 \\ \hline
\end{tabular}  
\caption{Parallel episodes for Example 3. 
The data is generated using (\ref{eq:lambda-update}) for updating firing rate of 
neurons}
\label{tab:pattern-2-parallel}
\end{table}

\begin{table}[!htb]
\centering
\begin{tabular}{|r|r|r|r|l|}
 \hline
Episode & Freq. & Time & Size & Patterns \\
expiry & Th. & (sec) & (No.) & Discovered \\
\hline
0.001 & 0.01 & 0.3 & 3(1) & \small{A B C : 492} \\ \hline
0.002 & 0.01 & 0.35 & 3(1) & \small{A B C : 505} \\ \hline
0.004 & 0.01 & 0.3 & 4(1) & \small{A B C D : 431} \\ \hline
0.006 & 0.01 & 0.32 & 4(2) & \small{X A B C : 449} \\
 &  &  &  & A B C D : 439 \\ \hline
\end{tabular}
\caption{Parallel episodes for Example 3. 
The data is generated using (\ref{eq:lambda-update2}) for updating firing rate of 
neurons}
\label{tab:pattern-2-parallel-2}
\end{table}

\begin{table}[!htb]
\centering
\begin{tabular}{|r|r|r|r|l|}
 \hline
Inter-event & Freq. & Time & Size & Patterns \\
interval & Th. & (sec) & (No.) & Discovered \\
\hline
0.000-0.002 & 0.01 & 0.32 & 2(6) & \small{A C : 385; B A : 376} \\
 &  &  &  & \small{B C : 373; A B : 372} \\
 &  &  &  & \small{C A : 361; C B : 355} \\ \hline
0.002-0.004 & 0.01 & 0.37 & 2(4) & \small{E F : 783; A D : 656} \\
 &  &  &  & \small{C D : 651; B D : 646} \\ \hline
0.004-0.006 & 0.01 & 0.28 & 2(3) & \small{X A : 790; X B : 774} \\
 &  &  &  & \small{X C : 769} \\ \hline
0.006-0.008 & 0.01 & 0.29 & 2(2) & \small{D E : 720; X D : 454} \\ \hline
\end{tabular}  
\caption{Serial Episodes for Example 3. 
The data is generated using (\ref{eq:lambda-update}) for updating firing rate of 
neurons}
\label{tab:pattern-2-serial}
\end{table}

\begin{table}[!htb]
\centering
\begin{tabular}{|r|r|r|r|l|}
 \hline
Inter-event & Freq. & Time & Size & Patterns \\
interval & Th. & (sec) & (No.) & Discovered \\
\hline
0.000-0.002 & 0.01 & 0.35 & 2(6) & \small{A C : 301; B A : 297} \\
 &  &  &  & \small{B C : 271; A B : 262} \\
 &  &  &  & \small{C A : 261; C B : 254} \\ \hline
0.002-0.004 & 0.01 & 0.41 & 2(4) & \small{E F : 584; A D : 566} \\
 &  &  &  & \small{C D : 551; B D : 562} \\ \hline
0.004-0.006 & 0.01 & 0.25 & 2(3) & \small{X A : 602; X B : 598} \\
 &  &  &  & \small{X C : 627} \\ \hline
0.006-0.008 & 0.01 & 0.35 & 2(2) & \small{D E : 554; X D : 202} \\ \hline
\end{tabular}
\caption{Serial Episodes for Example 3. 
The data is generated using (\ref{eq:lambda-update2}) for updating firing rate of 
neurons}
\label{tab:pattern-2-serial-2}
\end{table}

\begin{table}[!htb]
\centering
\begin{tabular}{|r|r|r|r|l|}
 \hline
Inter-event & Freq. & Time & Size & Patterns \\
interval & Th. & (sec) & (No.) & Discovered \\
\hline
\{0.000-0.002, & 0.01 & 1.37 & 5(1) &  \\
0.002-0.004, &  &  &  & \small{$X^{\underrightarrow{0.004-0.006}} [A B C]$}  \\
0.004-0.006, &  &  &  & \small{$^{\underrightarrow{0.002-0.004}} D^{\underrightarrow{0.006-0.008}}$} \\
0.006-0.008, &  &  &  & \small{$E^{\underrightarrow{0.002-0.004}} F$ : 372} \\
0.008-0.010\} &  &  &  &  \\ \hline
\end{tabular}  
\caption{Synfire episodes for Example 3. 
The data is generated using (\ref{eq:lambda-update}) for updating firing rate of 
neurons}
\label{tab:pattern-2-synfire}
\end{table}

\begin{table}[!htb]
\centering
\begin{tabular}{|r|r|r|r|l|}
 \hline
Inter-event & Freq. & Time & Size & Patterns \\
interval & Th. & (sec) & (No.) & Discovered \\
\hline
\{0.000-0.002, & 0.01 & 2.1 & 5(1) &  \\
0.002-0.004, &  &  &  & \small{$X^{\underrightarrow{0.004-0.006}} [A B C]$}  \\
0.004-0.006, &  &  &  & \small{$^{\underrightarrow{0.002-0.004}} D^{\underrightarrow{0.006-0.008}}$} \\
0.006-0.008, &  &  &  & \small{$E^{\underrightarrow{0.002-0.004}} F$ : 287} \\
0.008-0.010\} &  &  &  &  \\ \hline
\end{tabular}
\caption{Synfire episodes for Example 3. 
The data is generated using (\ref{eq:lambda-update2}) for updating firing rate of 
neurons}
\label{tab:pattern-2-synfire-2}
\end{table}

The results for parallel episode discovery (see Tables 
\ref{tab:pattern-2-parallel}  and \ref{tab:pattern-2-parallel-2}) 
show that $(ABC)$ is the group of
neurons that co-spike together. The serial episode discovery results
are given in Tables \ref{tab:pattern-2-serial} and 
\ref{tab:pattern-2-serial-2}. As can be seen from 
the table, with different 
 pre-specified inter-event time constraints we can discover only 
different parts of the underlying network graph because no single 
inter-event constraint captures the full pattern. 

As in Example 2, we replace occurrences of parallel episode with a 
new event in the data stream. 
We then run Algorithm~\ref{alg:count-serial-INTERVAL-discovery} to 
discover serial episodes along with inter-event constraints, given   
 a set of possible inter-event intervals. 
The results obtained are shown in 
Tables \ref{tab:pattern-2-synfire} and \ref{tab:pattern-2-synfire-2}. 
As can be seen from the table, we can discover correctly 
 the underlying network pattern. This example illustrates our algorithm for 
 choosing inter-event constraints automatically. Given a set of possible intervals, 
the algorithm has been able to correctly identify the different synaptic delays 
in the network pattern. As in the other two examples, we get the same results 
in terms of frequent episodes irrespective 
of the method used to update the firing rates of neurons.

The examples presented in this subsection illustrate the effectiveness of our temporal 
data mining method in inferring useful information regarding connectivity patterns. 
Here we have presented the case where only one connectivity 
pattern is embedded. We have tested the algorithm with multiple such patterns also. 
We have used up to four different patterns of varying lengths all embedded together and 
the algorithms are found to be effective in inferring all the connectivity patterns. 
(See \cite{Deb2006,archive-report} for details of these results). 

\subsubsection{Assessing significance of discovered patterns}

The examples above show that if we generate spike data using special embedded 
patterns in it then our algorithms can detect them. However, this does not answer the 
question: if the algorithm detects some frequent episodes what confidence do we have 
that they correspond to some patterns. 
This essentially concerns the statistical significance of the discovered episodes. 

To answer this question we have to essentially choose a {\em null hypothesis} that asserts
that there is no `structure' in the system generating the data. Then we need to
calculate the probability that an episode of a given size would have a given frequency 
(that is, number of non-overlapped occurrences) 
in the data generated by such a model and this will tell us what is the chance of a
discovered frequent episode coming up
by chance in `random' data and hence tells us the statistical
significance of the discovered frequent episodes. This can also allow us to calculate
the frequency threshold so that all frequent episodes (with frequency above this
threshold) are statistically significant at a given level of confidence. We are currently 
developing appropriate statistical models for such a null hypothesis to derive the 
required confidence bounds and would be discussing them in our future work. For this 
paper, we present empirical results to show that long episodes with high frequency do 
not come up unless the data contains specific biases for coordinated firing by neurons. 

Currently, in all reported methods of spike data analysis, whenever 
 issues of statistical significance are considered, it is 
always with a null hypothesis of {\em iid}
processes generating spike data.\footnote{One notable exception is 
\cite{amarasing-thesis}
where more complicated null hypotheses are considered. However, this work does not
deal with finding useful patterns in spike data; the objective of the
analysis there is to determine the time scale at which exact times of spikes may carry
useful information as opposed to all information being carried by only the spiking
frequency}. However, we would like a null hypothesis that says there is no 
strong structure in the interactions among neurons. 
Being able to reject a null hypothesis of {\em iid} processes generating data, does not 
seem appropriate when we want to assert that the patterns we found 
denote strong interactions among neurons. 
Hence, we want to consider a composite null hypothesis in which we
include  not only {\em iid} processes, but also other models
for interdependent neurons without any specific strong connectivity patterns or strong
predispositions for coordinated firing. Due to difficulties with analytical tractability, 
 instead of an analytical approach,  
we take the empirical approach of capturing our null hypothesis in
a simulation model and estimate the relevant probabilities by generating many random
data sets from such a model. (This approach is similar in spirit to the so called
`jitter' method \cite{date2001}).

We generate our random data sets using essentially three different types of models.
For the first one, we use the same simulator as described
earlier; but we allow only the random interconnections.  (The weights of
interconnections would be uniformly distributed over a range that represents small values for 
the conditional probability of the receiving neuron firing given that the sending neuron 
has generated a spike). 
Next, we
generate data sets by assuming that different neurons generate spikes as independent
Poisson processes by simply choosing random fixed rates for the neurons. In this, we
also include cases where many neurons can have the same firing rate. For this, we fix
five or ten different random firing rates and randomly assign each neuron to have
one of these firing rates. For our third type of data sets, we include
models where rates of firing by neurons change; but without any relation to firing by
other neurons. For this we choose random firing rates for neurons and  at
$\Delta T$ time steps we randomly change the firing rate. Here also we include the
case where firing rates of some random subsets of neurons are all tied together.

Thus our null hypothesis includes
models where different neurons could be {\em iid} Poisson processes, or
inhomogeneous Poisson processes where the firing rates may be correlated but the rate
is not dependent on firing of other neurons. In addition,
our null hypothesis also includes models with interdependent neurons 
where rates of firing change
based on spikes output by other neurons but without any bias for specific strong
interconnectivity patterns. We feel that this is a large enough set of models to
consider in the null hypothesis. If, based on our episode frequencies, we can
reject the null hypothesis, then, it provides reasonably strong evidence that episodes with
sufficiently high frequency can not come about unless there is a bias or interdependence
in the underlying neural system for coordinated firing by some groups of neurons.

The specific random data sets are as follows. All data sets are from
a 26 neuron system. We use six different types of random data to be
called Noise-1 to Noise-6. Each of these is explained below.
\begin{itemize}
\item[Noise-1]: We use an interconnected neuron system  but with
only random interconnections. Here we use the first method of updating firing rate
given by (\ref{eq:lambda-update}). The firing rate under zero input is chosen to be
20 Hz. The interconnection weights are uniformly distributed
in $[-0.75, \ 0.75]$ which corresponds to a conditional probability range of
$[0.012, \ 0.032]$. Since the normal firing rate is 20 Hz, the probability of
any neuron firing in an interval of 1 ms is about 0.02. If neurons are independent
then conditional probability of $B$ firing given $A$ fired earlier would still be
0.02. Thus the random interconnections change this conditional probability in either
direction by about 50\%.
\item[Noise-2]: Same as Noise-1 except that we use the linear model given by
(\ref{eq:lambda-update2}) for updating the firing rate based on input. The weights for
random interconnections are so chosen that they correspond to variations in the
conditional probability in the range of $[0.012, \ 0.032]$ as in Noise-1.
\item[Noise-3]: Here we take each of the neurons to be spiking as independent
Poisson processes with fixed firing rate chosen at random from the interval
$[10, \ 30]$Hz.
\item[Noise-4]: Same as Noise-3 but each Poisson process has time-varying rate.
At every $\Delta T$, each neuron chooses a firing rate at random using a uniform
distribution over $[10, \ 30]$.
\item[Noise-5]: Neurons are randomly made into five groups with all neurons
in a group having the same firing rate. The firing rate of each group is
fixed and is set at random from the interval $[10, \ 30]$.
\item[Noise-6]: Same as Noise-5 but the common firing rate of all neurons in a group
is made time varying by randomly changing it every $\Delta T$.
\end{itemize}

For each of these six types of noise data we generate 25 data sets. On each data set we 
mine for parallel episodes with usual expiry time constraint and serial episodes with usual 
inter-event time constraint, for various sizes of episodes. For each episode size $n$, we find 
the maximum frequency observed for any episode of that size in each data set and calculate 
the average of the maximum episode frequency over the 25 data sets. (In all the 
simulations described here, we use a frequency threshold of zero. While this significantly 
increases the computational burden, it is necessary to exactly determine maximum 
frequencies in noise data). 

The aim is to compare this maximum frequency of episodes of different sizes (in noise data) 
with the minimum 
observed frequency of episodes of that size when some patterns are embedded. 
For this we also generate data with embedded patterns using our simulator. 

For comparison 
in the case of parallel episodes, we embed a synchrony pattern of size 10 in the network.
We used the sigmoidal update rule for firing rate as given by 
(\ref{eq:lambda-update}) and fixed the weight values by specifying that the conditional 
probability needed is 0.8.  
We generate 25 data sets with the embedded pattern and mine for parallel episodes with 
expiry constraints. In each data set, for each size $n$, $n=1, \ldots, 10$, 
we find the minimum frequency among all parallel episodes of size $n$ which are part of the 
embedded pattern. We find average of this minimum frequency (for each $n$) over the 
25 data sets. 

For the case of serial episodes we embed a ordered firing pattern of size 10. As explained 
earlier, to embed such patterns, we calculate the weight of interconnections between 
successive neurons by specifying the desired conditional probability, $\rho$. 
For this comparison, we 
use different values for this conditional probability. Using the sigmoidal update rule  
 given by (\ref{eq:lambda-update}), we generate 25 data sets each with the 
conditional probability being 0.8, 0.6 and 0.4. We also generated 25 data sets each using the 
linear update model given by (\ref{eq:lambda-update2}) with the conditional probability 
values being 0.8 and 0.7. In each data set, we mine for serial episodes with inter-event time 
constraint. For each size $n$, $n=1, \ldots, 10$, we find the minimum frequency of an 
episode of size $n$ among all episodes of that are part of the embedded serial pattern. 
We then find the average of the minimum episode frequency (for size $n$) over the twenty five 
data sets. This we do for each of the five different methods of embedding the serial 
pattern as explained above.  

\begin{figure}
  \centerline{
     \epsfig{file=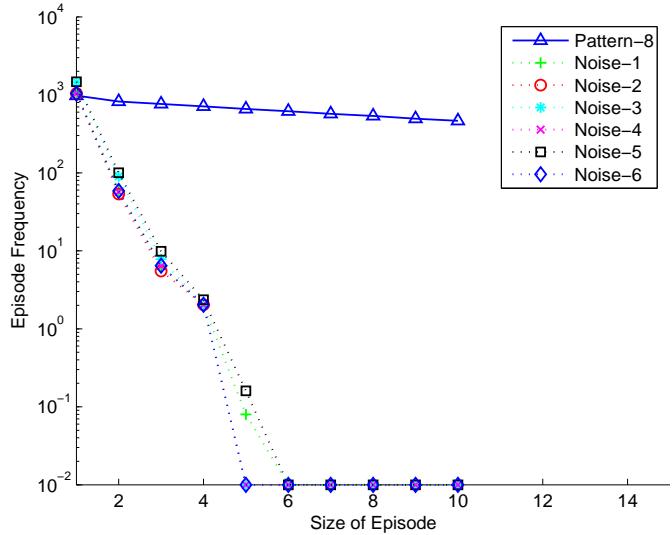, width=4.0in}
     }
  \caption{Statistical significance results for parallel episodes. Comparison of 
\textbf{maximum} frequencies of parallel episodes of different sizes for data with no structure 
and and \textbf{minimum} frequencies for data containing embedded patterns. Noise-1 to Noise-6 represent different types of 
data with no structure in it as explained in text. Pattern-8 refers to data generated 
with a model containing a synchrony pattern of size 10}
  \label{fig:stat-par-comp}
\end{figure}

\begin{figure}
  \centerline{
     \epsfig{file=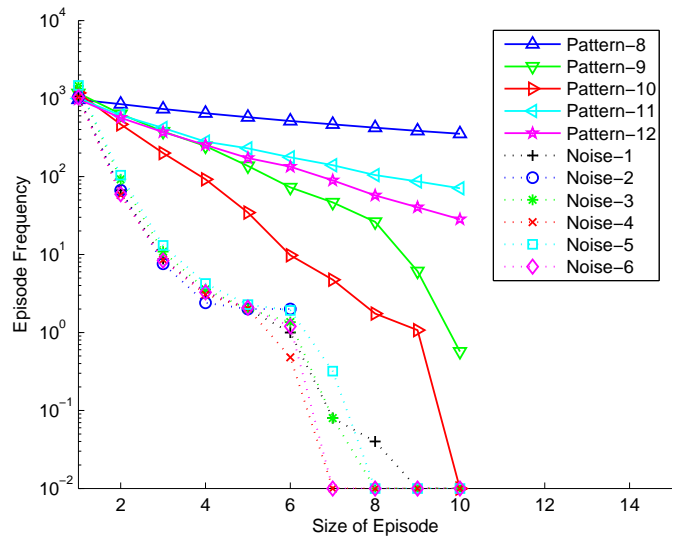, width=4.0in}
     }
  \caption{Statistical significance results for serial episodes. Comparison of 
\textbf{maximum} frequencies of serial episodes of different sizes for data with no structure 
and \textbf{minimum} frequencies for data containing embedded patterns. 
Noise-1 to Noise-6 represent different types of 
data with no structure in it as explained in text. Pattern-8 to Pattern-12 refer to 
 to data generated with a model containing a ordered pattern of size 10. These different 
data sets differ in the strength of interconnection between successive neurons in the 
serial pattern embedded. Pattern-8, Pattern-9 and Pattern-10 contain connections which 
correspond to the conditional probability value for firing of the next neuron as 
0.8, 0.6 and 0.4 respectively. All these are obtained using the sigmoidal function 
for updating rate of firing. Pattern-11 and Pattern-12 are with connection strengths 
that correspond to conditional probability values of 0.8 and 0.7 respectively and 
these are generated using the linear model for updating the rate of firing.}
  \label{fig:stat-ser-comp}
\end{figure}

The results for parallel episodes are shown in fig.~\ref{fig:stat-par-comp} and those for 
serial episodes are shown in fig.~\ref{fig:stat-ser-comp}. Note that in both figures 
we have used a log scale on the Y-axis. The figures show how average frequency varies 
with size in each type of data sets. {\em As explained above, for the case of noise data, 
namely Noise-1 to Noise-7, what is plotted is average of {\bf maximum} 
observed episode frequency 
versus size of episode. For case of pattern data, what is plotted is the average of 
{\bf minimum} frequency of episodes that are part of the embedded pattern}. 

From fig.~\ref{fig:stat-par-comp}, it is clear that frequencies of parallel episodes that are 
part of the embedded patterns are higher by two orders of magnitude compared 
to those in noise data even at size 3. Also, frequencies of parallel episodes in noise 
data rapidly fall to zero with size. However, up to the size of embedded pattern, frequencies 
of parallel episodes of all sizes that are part of the embedded pattern
remain very high. This clearly demonstrates that you can not get such episodes of large 
size with appreciable frequency unless the underlying data generation model has 
biases for the required type of coordinated firing. 

From fig.~\ref{fig:stat-ser-comp}, similar trends are seen for serial episodes also. 
The frequencies in case of noise data fall to zero quickly with increasing size of 
episodes and are also orders of magnitude smaller than those for data with patterns. 
In addition, in this figure we can see a clear empirical demonstration of our 
intuitive idea of why data mining can uncover the strong connections. Here we have used 
five different types of data with serial patterns with different data having interconnections 
of different strengths. As explained earlier, the weight of interconnection can also be 
equivalently specified in terms of the conditional probability, $\rho$. We show here 
cases of pattern data where the conditional probabilities vary from 0.4 to 0.8. At high 
conditional probability, all episode frequencies remain high up to size 10. But as the 
conditional probability is decreased, the frequencies come down. For example, when the 
conditional probability is only 0.4, the frequencies of episodes come down appreciably with 
size of episodes (even though they are still higher than those for noise data sets). Looking 
at the curves corresponding to the five cases of data with patterns, we see that decrease 
of episode frequency with size is directly related to how strong are the connections among 
neurons in the embedded patterns. 
Thus, we can say that long episodes with high frequencies 
can not come about unless there are 
{\em strong} interconnections (represented by the episodes) in the underlying data generation 
model. 

The results presented in this subsection clearly demonstrate that when we find episodes 
with high frequencies they are statistically significant. Since the difference between 
maximum frequencies in Noise data and minimum frequencies in data with patterns is very 
high, it is fairly easy to select a frequency threshold. 
If we choose the same frequency thresholds as in our earlier examples on synthetic data, 
from the above figures, we see that there would be no frequent episodes under any of the noisy 
data. Since the average frequency  of episodes under our 
null hypothesis (i.e., noisy data)  are 
obtained from 150 samples (six noise models, each with 25 repetitions), we can 
say that the empirical probability of getting a frequent episode under our null 
hypothesis is less than $\frac{1}{150}$. Thus, (under this empirical estimate of 
probabilities) we can claim that the discovered frequent episodes are significant 
with a p-value of better than 0.006.

\subsection{\label{sec:Analysis-of-multi-neuron}Analysis of multi-neuron data obtained 
from MEA experiments}

In this subsection we present some of the results we obtained with our algorithms
on multi-neuronal data obtained through multi-electrode array experiments.\footnote{We
are grateful to Prof. Steve Potter, Georgia Institute of Technology and Emory
University, Atlanta, USA, for providing the data and for many useful discussions on
analyzing this data.} The data is obtained from dissociated cultures of cortical
neurons grown on multi-electrode arrays. This is an extremely rich set of data where
58 cultures of varying densities are followed for five weeks. Everyday, the spontaneous
activity as well as
 stimulated activity of each culture is recorded for different time durations.
(See \cite{Potter2006} for the details of experiments, nature of data, trends observed etc.).
Since data was recorded from each culture for many days, one can presumably infer development
of connections also. Here we only present a few of the results we obtained from analyzing the
spontaneous data from these
cultures, to illustrate the utility of our temporal data mining techniques on MEA data.

This data is not spike sorted and detected spikes are only assigned to the electrode 
on which the signal is recorded. Hence, in our data mining algorithm, the event types 
would be electrode names and not neurons. All our episodes only indicate connectivity 
patterns involving electrodes. However, by taking each occurrence of a 
detected frequent episode and looking at the raw signal at those times, it is possible 
to infer whether or not same neuron is involved in the spikes on that electrode. For the 
purposes of illustration here, we only discuss episodes in terms of electrodes. We 
note that such information is also very useful for making reasonable hypotheses 
regarding underlying connectivity patterns.  

In these dissociated cortical cultures, there is a lot of spontaneous activity including
many cycles of network-wide bursts \cite{Potter2006}. Thus, patterns of coordinated
firing by groups of neurons due to synapses,  would be rare in the sense that the
spikes which form the coordinated activity constitute only
 a small fraction of the total number of
spikes output by the system. Thus, simple cross correlation based methods are not very
effective in unearthing coordinated firing patterns. Using our algorithm for serial
episode discovery under inter-event constraints, we are able to obtain some frequent
episodes which remain frequent for a large number of days with increasing trend in
frequency.

\begin{figure}[!htb]
\centering
\epsfig{file=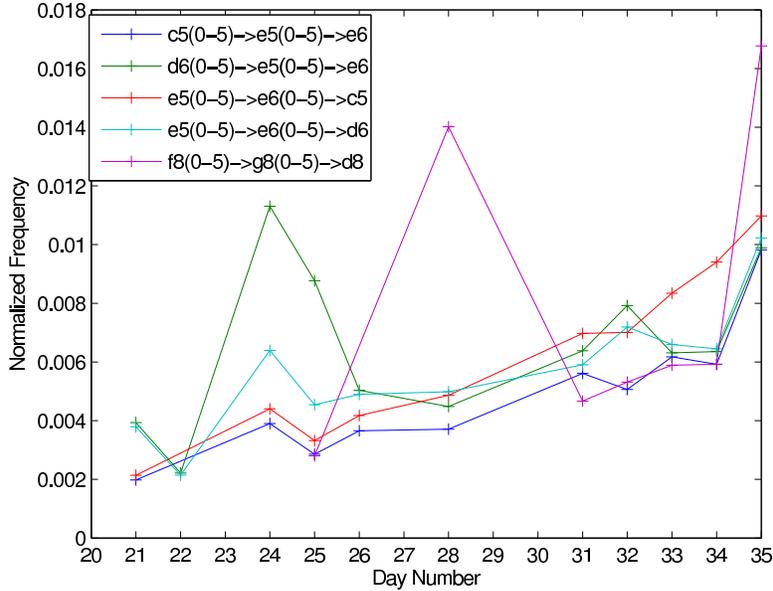, width=4.0in}
\caption{Some frequent Serial Episodes discovered from multi-electrode array data. 
We plot the normalized frequency versus the age in days of the
culture. These are from data of culture 2-1 of Wagenaar, Pine, and Potter (2006)\cite{Potter2006}.}
\label{fig:potter-norm-freq}
\end{figure}

Fig.~\ref{fig:potter-norm-freq} shows a few such serial episodes discovered
in the data from one of the cultures. (We have used inter-event time constraint of
1--5 ms). The figure plots the frequency (in terms of the
number of non-overlapping occurrences as a fraction of the data length) for the frequent
serial episodes versus the day on which the data is collected. In the figure, c5, e5, e6,
d6 etc. are the pin numbers (or electrode names) 
 in the multi-electrode array which will be event types for the
data mining algorithms. The increasing trend of the frequency is very clear and it is
highly plausible that these episodes represent some underlying microcircuits that are
developing as the culture ages. 

In this data, there is no ground truth available regarding connections and hence it
is not possible to directly validate the discovered episodes. However, we can indirectly
get some evidence that the episodes capture some underlying structure in the neural system
by looking at the sets of episodes obtained from same culture on different days and from
different cultures. We considered six cultures, namely, culture 2-1 to culture 2-6. For
each culture we considered the data from the last five days, namely days 31 to 35. (As we
have seen from the earlier figure, the circuits seem to stabilize only in the last week).
However, in our data set, for culture 2-4 there was no recording on day 34
Thus we have 29 data sets such as 2-1-31 (meaning culture 2-1, day 31) and so on.
From each culture on each day, we have 30 minutes of data recording spontaneous activity.
From each data
set, we have taken a 10 minute duration data slice.
From each such data
slice, we identified top twenty most frequent 7-node serial episodes with inter-event
interval constraint of 1--5 ms. (We want to consider long episodes because, as we
saw earlier, it is highly unlikely to have large size frequent episodes by chance. The
size of 7 is chosen so that all data sets have at least twenty episodes of that size). Now
we want to compare the sets of episodes discovered from different data slices. For this we
need a measure of similarity between sets of episodes.

We define a similarity score for two sets, $A, B$,  of episodes of size, say, N, as follows.
We first count the
number of N-node episodes that are common in the two sets and remove all the common
ones from both sets. Then we replace each episode (in each set) with the two (N-1)-node
subepisodes  obtained by dropping the first or last nodes in the original episode. We now
count the common (N-1)-node episodes (in the two sets) and remove them. We go on like this,
by replacing the left-over episodes with subepisodes of size one less and counting the
common ones, till we reach episodes of size 1. Let $n_i$ denote the number of common
episodes of size $i$. Then the similarity between the sets $A$ and $B$ is defined as
\[\mbox{Sim}(A, B) = \sum_{i=1}^N \; 2^i n_i.\]
Since we want to view episodes as representing connections, similarity has to capture
how much of the paths represented by different episodes are common. The above measure
does just that and gives higher weightage to common long episodes. We would like to point out
that this particular similarity score is rather arbitrary and is used only for illustration.

\begin{figure}[!htb]
\centering
\epsfig{file=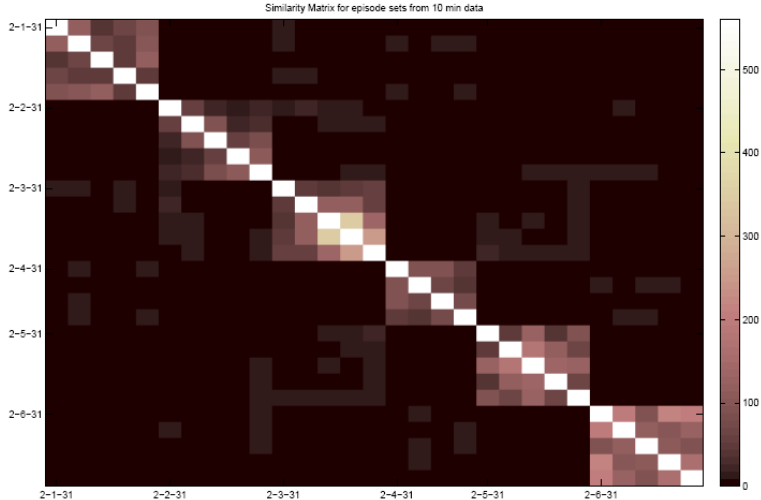, width=4.0in}
\caption{Cross similarity between sets of frequent Serial Episodes. Each pixel  
represents the similarity score between two sets of frequent episodes. 
Each axis consists of groups of (culture-dish) combinations going from 2-1 to 2-6.
In each group, 5 days (i.e. day 31 to day 35) of recording are shown with label given to the start day (e.g. 2-1-31).
The reverse diagonal corresponds to the case where the two sets of frequent episodes are identical.
The similarity values are represented by different gray levels 
as  explained by the legend on the
right. Note the high similarity scores around the reverse diagonal. It shows
that sets of episodes from the same culture have higher similarity than those from
different cultures.}
\label{fig:potter-sim-matrix1}
\end{figure}

Fig.~\ref{fig:potter-sim-matrix1} shows the cross-similarity between the
 sets of frequent episodes from the 29 data slices
by color coding similarity values. The axes indicate the culture-day
combinations.  Note that the two axes are
ordered differently so that the reverse diagonal represents similarity between
identical sets of episodes. That is why the reverse diagonal has highest
similarity values.  What is
interesting is that data slices from the same culture but from different days are
highly similar. This can be seen by observing the $5 \times 5$ submatrices
around the reverse diagonal in the figure. (For the 2-4 culture, this submatrix
is only $4 \times 4$ because there is no data for 2-4-34).
This is in sharp contrast to the fact (as seen from the figure) that
sets of episodes obtained from different cultures have very low similarity. 

That the set of frequent episodes characterizes a culture like this, is interesting. 
In such cultures, qualitatively the spiking behavior observed in all cultures is very 
similar \cite{Potter2006}. Since these cultures are seeded by the same cortical neurons 
and grow with time, they most likely form random interconnection patterns. Thus, the 
interconnectivity pattern may be the only thing that distinguishes one culture from 
another. Thus, our results seem to point out that the frequent episodes capture, in some 
sense, the underlying connectivity pattern. We feel that our results on synthetic data 
along with the results presented in this subsection provide sufficient justification 
that our temporal data mining techniques constitute a promising approach for analyzing 
multi-neuronal spike train data.

\section{Discussion}
\label{sec:conc}

Analyzing multi-neuron spike data is a challenging problem of much current interest in
neuroscience. With the recent advances in experimental techniques, 
we can now easily obtain data
representing the simultaneous activity of hundreds of neurons. Hence algorithms that
can discover significant patterns of co-ordinated spiking activity among neurons 
would be very useful in making sense of these vast amounts of data. Discovering such patterns 
would help in understanding the underlying connectivity structure in the neural tissue 
and to relate it to the function of the nervous system. 
Such an understanding
of the behavior of interacting neurons is very useful in elucidating issues
such as learning and memory as well as for applications such as brain computer
interfaces.

The main objective of this paper is to show the utility of a class of temporal data mining 
techniques for the above. We have introduced the notion of 
mining for frequent episodes with temporal constraints and have presented  
algorithms for finding such frequent episodes from large data streams. 
We have shown how one can detect many coordinated firing patterns 
in multi-neuronal spike data, such as order,
synchrony, and synfire chains in terms of episodes with appropriate temporal
constraints. We illustrated the effectiveness of the algorithms by analyzing synthetically
generated spike sequences that have embedded patterns in them. For this we have modeled
each neuron as an inhomogeneous Poisson process whose spiking rate gets modified in
response to the input received from other neurons. By building an interconnected  system
of such neurons with some specific large excitatory connections along with many small random
connections, we can embed different patterns in the system that is generating the spike
data. Since the ground truth is known in these sequences, they serve as useful test
beds for assessing the capabilities of our algorithms. We have shown that our algorithms
unerringly discover the underlying connectivity structure. We have also shown that our 
algorithms can properly choose the inter-event temporal constraints by using 
data generated by networks where different synapses have different delays. 

We have also used our neuronal spike data simulator to study the statistical significance 
of discovered frequent episodes. We have generated many sets of random data
of both independent and dependent neurons but without any strong synapses. 
It is seen that the maximum
frequency of episodes in such noise data are orders of magnitude smaller than minimum
frequencies of relevant episodes when data contains patterns. Also the frequencies 
of episodes fall off very rapidly with the size (length) of the episodes 
in noise data when compared with that 
in data generated using some strong interconnections among neurons. These empirical results 
provide strong evidence that long episodes with appreciable frequencies can not 
come up by chance if the underlying system does not have the needed connectivity 
pattern. This, we feel, is a very
significant contribution of this paper because the null hypothesis we consider here
goes beyond the usual one of independent homogeneous Poisson processes. 
We have also provided some evidence for the effectiveness of these methods by analyzing 
data obtained from cortical neurons cultured on multi electrode arrays. 

The data mining techniques discussed here are very efficient computationally. 
In addition to this, data mining techniques are attractive because such algorithms are  
model independent in the sense that they do not need to assume anything regarding 
a model for the interacting neurons. We have used two different methods of updating the 
firing rate of a neuron in response to inputs received from others. These 
result in different spike trains and the number of occurrences of different 
episodes are also different. However, the frequent episode discovery method is 
very good at inferring the underlying connectivity structure irrespective of the 
model followed for data generation.

The algorithms and results presented here should be viewed as indicative of the potential of 
temporal data mining for this problem. 
An objective of this paper is to introduce the problem 
multi-neuron spike data analysis to data mining community. Though this is a 
very challenging problem of analyzing large data sets to find underlying 
patterns, there does not seem to be much work in exploring data mining 
techniques for this problem. 
One can think of this problem as one of unearthing the network connectivity 
pattern given only node-level dynamic data. Such a problem would be relevant 
in many other application areas as well. For example, analyzing abnormal 
behavior of communication networks, finding hidden causative chains in 
complex manufacturing processes controlled by distributed controllers, etc. 
We hope our work presented here would contribute toward developing of 
data mining techniques relevant in such applications as well.

There are many open issues in the methods 
presented here. 
While the serial episodes with inter-event constraints give us a 
good idea of connectivity pattern, they do not allow us to infer the exact connectivity 
pattern or the graphical structure. For example, 
consider a circuit where $A$ is connected to $B$ with a synapse 
of delay 5 ms and $A$ is also connected to $C$ with a synapse of delay 10 ms. Hence, 
every time $A$ spikes, $B$ will spike 5ms later (with a large probability) and $C$ 
will spike another 5ms later. Thus serial episodes $A\rightarrow B$, $B \rightarrow C$, 
$A \rightarrow B \rightarrow C$ with inter-event time of 5ms and serial episode 
$A \rightarrow C$ with inter-event time of 10ms would all be frequent. Now suppose 
there is also  a synapse between $B$ and $C$ of 5ms delay. Even then, the same set of 
episodes would be frequent. Thus the episodes alone can not completely determine 
the connectivity structure.  In this example, the difference would be that, in one case 
every occurrence of the $B \rightarrow C$ episode would be part of an occurrence 
of $A \rightarrow B \rightarrow C$ episode while in the other case it is not. Thus, 
more algorithmic as well as statistical techniques are needed to infer the circuits 
completely given the episodes. 
Another direction in which the work of this paper can be 
extended is in developing a proper analytical hypothesis testing framework for 
assessing statistical significance of the episodes with a null hypothesis that allows for 
weak interactions among neurons. 
The techniques presented in this paper are 
useful in unearthing information regarding only those connectivity patterns which are 
due to excitatory synapses because the search is for frequently occurring episodes. 
The neural circuits contain both excitatory as well as 
inhibitory synapses. Developing data mining methods to infer inhibitory connections 
would be another major extension of this work. We would be addressing these issues in 
our future work.

Analyzing multi-neuronal spike data to finally obtain useful information 
about the underlying functional connectivity is a challenging problem. 
It would need an interdisciplinary approach and concerted effort by many 
researchers to solve this problem to any reasonable level.  
From signal-processing techniques for detection \&  temporal localization 
of spiking events to analytical techniques for 
properly understanding their statistics, many different methods and algorithms are needed 
here. The point we wish to make is that the field of data mining  
has an important role to play. We hope that our paper contributes toward 
development of novel data mining techniques useful in this endeavor. \\

\begin{center}
{\bf Acknowledgments}
\end{center}

We would like to thank Mr. Arvind Murthy for all the results 
presented in Sec.~\ref{sec:Analysis-of-multi-neuron}. This work 
is partially supported by project funding from General Motors 
R\&D Center through SID, IISc.

\pagebreak

\appendix
\appendixpage
\section{Pseudo-code listing for Algorithms in the paper}

\begin{algorithm}[!htb]
\caption{\label{alg:count-parallel-EXPIRY}Non-overlapped count for parallel
episodes with expiry time constraint}
\begin{algorithmic}[1]
\REQUIRE Set $C$ of candidate $N$-node parallel episodes, event
streams $s=\langle(E_{1},t_{1}),\ldots,(E_{n},t_{n})\rangle$, frequency
threshold $\lambda_{min}\in[0,1]$, expiry time $T_{x}$
\ENSURE The set $F$ of frequent serial episodes in $C$

\FORALL {event types $A$}
	\STATE $waits(A)=\phi$
\ENDFOR

\FORALL {$\alpha\in C$}
	\STATE Initialize $autos(\alpha)=\phi$
\ENDFOR

\FORALL {$\alpha\in C$}
	\FORALL {event types $A\in\alpha$}
		\STATE Create node $s$ with $s.episode=\alpha$; $s.init=0$ ;
		\STATE $s.count=1$
		\STATE Add $s$ to $waits(A)$ 
		\STATE Add $s$ to $autos(\alpha)$
	\ENDFOR
	\STATE Set $\alpha.freq=0$
	\STATE Set $\alpha.counter=0$
\ENDFOR

\FOR {$i=1$ to $n$}
	\FORALL {$s\in waits(E_{i})$}
		\STATE Set $\alpha=s.episode$
		\STATE Set $j=s.count$ 
		\IF {$j>0$}
			\STATE Set $s.count=j-1$
			\STATE $\alpha.counter=\alpha.counter+1$
		\ENDIF
		\STATE $s.init=t_{i}$
		\STATE \{Expiry check\}
		\IF {$\alpha.counter=N$}
			\FORALL {$q\in autos(\alpha)$}
				\IF {$(t_{i}-q.init)>T_{x}$}
					\STATE $\alpha.counter=\alpha.counter-1$
					\STATE $q.count=q.count+1$
				\ENDIF
			\ENDFOR
		\ENDIF
		\STATE \{Update episode count\}
		\IF {$\alpha.counter=N$}
			\STATE Update $\alpha.freq=\alpha.freq+1$
			\STATE Reset $\alpha.counter=0$
			\FORALL {$q\in autos(\alpha)$}
				\STATE Update $q.count=1$
			\ENDFOR
		\ENDIF
	\ENDFOR
\ENDFOR
\STATE Output $F=\{\alpha\in C$ such that $\alpha.freq\ge n\lambda_{min}\}$
\end{algorithmic}
\end{algorithm}

\begin{algorithm}[!htb]
\caption{\label{alg:count-serial-INTERVAL-discovery}\small{Non-overlapped serial
episodes count with inter-event interval constraints}}
\begin{algorithmic}[1]
\small{
\REQUIRE Set $C$ of candidate $N$-node parallel episodes, event
streams $s=\langle(E_{1},t_{1}),\ldots,(E_{n},t_{n})\rangle$, frequency
threshold $\lambda_{min}\in[0,1]$, expiry time $T_{X}$
\ENSURE The set $F$ of frequent serial episodes in $C$
\FORALL{event types $A$}
	\STATE Initialize $waits(A)=\phi$
\ENDFOR
\FORALL{$\alpha\in C$}
	\STATE Set $prev=\phi$
	\FOR{$i=1$ to $N$} 
		\STATE Create $node$ with $node.visited=false$;
		$node.episode=\alpha$;
		$node.index=i$;
		$node.prev=prev$;
		$node.next=\phi$
		\IF{$i=1$}
			\STATE Add $node$ to $waits(\alpha[1])$ 
		\ENDIF
		\IF{$prev\ne\phi$}
			\STATE $prev.next=node$
		\ENDIF
	\ENDFOR
\ENDFOR
\FOR{$i=1$ to $n$}
	\FORALL{$node\in waits(E_{i})$}
		\STATE Set $accepted=false$
		\STATE Set $\alpha=node.episode$
		\STATE Set $j=node.index$ 
		\STATE Set $tlist=node.tlist$
		\IF{$j<N$}
			\FORALL{$tval\in tlist$}
				\IF{$(t_{i}-tval.init)>\alpha.t_{high}[j]$}
					\STATE Remove $tval$ from $tlist$
				\ENDIF
			\ENDFOR
		\ENDIF
		\IF{$j=1$}
			\STATE Update $accepted=true$
			\STATE Update $tval.init=t_{i}$
			\STATE Add $tval$ to $tlist$
			\IF{$node.visited=false$}
				\STATE Update $node.visited=true$
				\STATE Add $node.next$ to $waits(\alpha[j+1])$ 
			\ENDIF
		\ELSE
			\FORALL{$prev\_tval\in node.prev.tlist$}
				\IF{$t_{i}-prev\_tval\in(\alpha.t_{low}[j-1],\alpha.t_{high}[j-1]]$}
					\STATE Update $accepted=true$
					\STATE Update $tval.init=t_{i}$
					\STATE Add $tval$ to $tlist$
					\IF{$node.visited=false$}
						\STATE Update $node.visited=true$
						\IF{$node.index\le N-1$}
							\STATE Add $node.next$ to $waits(\alpha[j+1])$
						\ENDIF
					\ENDIF
				\ELSE
					\IF{$t_{i}-prev\_tval>\alpha.t_{high}[j-1]$}
						\STATE Remove $prev\_tval$ from $node.prev.tlist$
					\ENDIF
				\ENDIF
			\ENDFOR
		\ENDIF
		\IF{$accepted=true$ and $node.index=N$}
			\STATE Update $\alpha.freq=\alpha.freq+1$
			\STATE Set $temp=node$
			\WHILE{$temp\ne\phi$}
				\STATE Update $temp.visited=false$
				\IF{$temp.index\ne1$}
					\STATE Remove $temp$ from $waits(\alpha[temp.index])$
				\ENDIF
				\STATE Update $temp=temp.next$
			\ENDWHILE
		\ENDIF
	\ENDFOR
\ENDFOR
\STATE Output $F=\{\alpha\in C$ such that $\alpha.freq\ge n\lambda_{min}\}$
}
\end{algorithmic}
\end{algorithm}

\end{document}